\documentclass[12pt,preprint]{aastex}

\usepackage{graphicx}
\usepackage[english]{babel}
\usepackage[utf8]{inputenc}
\usepackage{comment}
\shortauthors{Karnath et al.}

\begin{document}

\shorttitle{The Dynamics, Structure, and Fate of a Young Cluster During Gas Dispersal}
\title{The Dynamics, Structure, and Fate of a Young Cluster During Gas Dispersal: Hectoschelle, Chandra, Spitzer, and Gaia Observations of CepOB3b}  

\author{
N. Karnath,\altaffilmark{1}
J.~K. Prchlik,\altaffilmark{2} 
R.~A. Gutermuth,\altaffilmark{3}
T.~S. Allen,\altaffilmark{4,5}
S.~T. Megeath,\altaffilmark{1}
J.~L. Pipher,\altaffilmark{6}
S. Wolk,\altaffilmark{2}
R.~D. Jeffries \altaffilmark{7}
}

\altaffiltext{1}{Ritter Astrophysical Research Center, Department of Physics and Astronomy, University of Toledo, Toledo, OH 43606, USA; nicole.karnath@rockets.utoledo.edu, s.megeath@utoledo.edu}
\altaffiltext{2}{Harvard-Smithsonian Center for Astrophysics, 60 Garden St., Cambridge, MA 02138, USA}
\altaffiltext{3}{Department of Astronomy, University of Massachusetts, Amherst, MA 01003, USA}
\altaffiltext{4}{Portland State University 1825 SW Broadway Portland, OR 97207, USA}
\altaffiltext{5}{Lowell Observatory, 1400 West Mars Hill Road, Flagstaff, AZ 86001, USA}
\altaffiltext{6}{Department of Physics and Astronomy, University of Rochester, Rochester, NY 14627, USA}
\altaffiltext{7}{Astrophysics Group, Keele University, Keele, Staffordshire ST5 5BG, UK}

\begin{abstract}

We present a study of the kinematics and structure of the Cep OB3b cluster based on new spectra obtained with the Hectoschelle spectrograph on the MMT and data from {\it Spitzer}, {\it Chandra}, and {\it Gaia}. At a distance of 819$\pm$16 pc, Cep OB3b is one of the closest examples of a young ($\sim$3 - 5 Myr), large ($\sim$3000 total members) cluster at the late stages of gas dispersal. The cluster is broken into two sub-clusters surrounded by a lower density halo. We fit the empirical density law of King (1962) to each sub-cluster to constrain their sizes and structure. The richer eastern sub-cluster has circular symmetry, a modest central density, and lacks molecular gas toward its core suggesting it has undergone expansion due to gas dispersal. In contrast, the western sub-cluster deviates from circular symmetry, has a smaller core size, and contains significant molecular gas near its core, suggesting that it is in an earlier phase of gas dispersal. We present posterior probability distributions for the velocity dispersions from the Hectoschelle spectra. The east will continue to expand and likely form a bound cluster with $\sim$35\% of stars remaining. The west is undergoing slower gas dispersal and will potentially form a bound cluster with $\sim$75\% of stars remaining. If the halo dissipates, this will leave two independent clusters with $\sim$300 members; proper motions suggest that the two sub-clusters are not bound to each other.

\end{abstract}

\keywords{stars: formation, pre-main sequence, kinematics and dynamics; open clusters and associations: Cep OB3b; techniques: radial velocities}

\section{Introduction}

Most stars form in embedded clusters \citep{carpenter2000,lada2003,megeath2016} and within a few Myr the natal gas will be expelled from the cluster relieving it of internal extinction assuming dust moves with the gas. Given that star formation efficiency is low, 4\% for entire clouds and 20\% in clusters \citep{lada2003,allen2007,evans2009,gutermuth2011a,megeath2016}, most of the mass leaves the cluster in a few Myr drastically lowering the gravitational potential of the cluster. The ensuing dynamical evolution will determine whether the cluster - in whole or in part - will form a bound open cluster or disperse into the galactic disk. Cep OB3b is an excellent environment to study this crucial step of evolution because at $\sim$3-5 Myr \citep{littlefair2010,allen2012}, the cluster has dispersed most of its natal gas and is observable at visible wavelengths. This gives a snapshot of cluster evolution toward the end of gas dispersal, in a birth cluster similar to that of the Sun \citep{adams2001}, and at a later stage of evolution than the Orion Nebula Cluster (ONC).

Many numerical studies have been carried out to determine how clusters evolve during and after gas dispersal using a range of assumptions for the initial cluster properties and the timescale for gas dispersal \citep[e.g.,][]{lada1984,adams2000,geyer2001,boily2003,baumgardt2007,chen2009,proszkow2009,goodwin2009,moeckel2010,pelupessy2012,farias2015,farias2018}. Observational studies are limited at this crucial time during gas dispersal of cluster evolution. Most of the large, young clusters within 1 kpc, in particular Orion, NGC 2264 or Mon R2, are partially embedded and appear to be in earlier stages of their gas dispersal \citep{dahm2005,gutermuth2011a}. Comparisons of the number of embedded and bound clusters within $\sim$2 kpc of the Sun indicate 7\% of embedded clusters survive gas expulsion to form open clusters \citep{lada2003}. Determining the structure and kinematics of clusters undergoing gas dispersal is a key step toward improving our understanding of what factors determine whether a cluster survives. 

\citet{allen2012} carried out a census of the young stellar objects (YSOs) in Cep OB3b. They employed $Spitzer$ data to identify stars with infrared excesses due to dusty disks or envelopes, a combination of new and archival $Chandra$ data to detect X-ray emission from coronae of young stars, and visible light photometry from the literature to identify stars on the Cep OB3b isochrone. \citet{allen2012} estimate there are $\sim$3000 total members in Cep OB3b consisting of primarily low-mass ($\le$ 1 M$_{\odot}$) stars. The density of stars is considerably less than the ONC \citep{hillenbrand1998,megeath2016} but it is comparable in size and membership to the ONC \citep{carpenter2000,allen2007}, making Cep OB3b one of the largest known, young clusters within 1 kpc of the Sun \citep{allen2012}. Most of the members, however, lie in a cavity of low extinction, implying that most of the natal gas has been dispersed. This combination of size and evolutionary state makes it an excellent region to study the effect of gas dispersal on young clusters. The spatial distribution of the objects shows a hierarchical morphology composed of two distinct sub-clusters, denoted east (eastern sub-cluster) and west (western sub-cluster), each associated with a distinct molecular clump within the larger Cep OB3 molecular cloud \citep{sargent1977,heyer1996}. Furthermore, a substantial difference in the disk fraction of the sub-clusters, 32$\pm$4$\%$ for the east and 50$\pm$6$\%$ for the west, was found by \citet{allen2012} and was attributed to a difference in the typical ages of the constituent stars (east being older) in the sub-clusters, rather than photoevaporation of disks by high mass members. These results suggest a distinct origin for both sub-clusters;  however, since the two sub-clusters share a contiguous region of high stellar surface density \citep{gutermuth2011b,allen2012} within the Cep OB3b cloud with a diffuse halo of stars surrounding it, we consider these as parts of a single cluster. Cep OB3b has one O star (O7V, HD 217086) that resides in the eastern sub-cluster and several B stars throughout the entire cluster \citep{blaauw1959}. 

In order to relate young cluster populations to older stars in the field and open clusters, we need to observe young clusters at this critical stage of their evolution. The goal of this paper is to assess the kinetic and potential energy of the two sub-clusters, determine the fates as bound clusters, and study the effect of gas dispersal on young clusters. We carry out a radial velocity (RV) survey of 499 stars in Cep OB3b to measure the velocity dispersion and measure the kinetic energy. This study is complemented by an analysis of the structure of the two sub-clusters using $Spitzer$ and $Chandra$ data; from these data we can determine the current potential energy of the cluster. Prior to this study, the structure and kinematics of Cep OB3b have not been assessed using the population of low mass stars. Finally, we use Gaia DR2 to both refine the distance to the cluster, which is needed to measure the potential energy of the cluster, and to measure the bulk motions of the sub-clusters. 

Velocity dispersions of YSOs have been measured for a number of young clusters and molecular clouds probing the kinematical states. The ONC has been observed several times in RV surveys to determine velocity dispersions: \citep[$\sim$1.8 km s$^{-1}$,][]{sicilia-aguilar2005}, \citep[$\sim$2.3 km s$^{-1}$,][]{furesz2008}, \citep[$\sim$2.5 km s$^{-1}$,][]{tobin2009}, \citep[$\sim$2.3 km s$^{-1}$,][]{kounkel2016}, and \citep[$\sim$1.7 km s$^{-1}$,][]{dario2017}. \citet{dario2014} found that the ONC is an expanding cluster undergoing gas expulsion, which is in agreement with \citet{kuhn2018} who measured a parallax-based velocity dispersion from Gaia DR2 data of $\sim$1.8 km s$^{-1}$.

Outside of Orion, Hectoschelle observations of NGC 2264 by \citet{furesz2006} \& \citet{tobin2015} found a one-dimensional velocity dispersion of $\sim$2.5 km s$^{-1}$ and classified the region as several clumps of star formation and not one bound cluster. NGC 1333 was studied with APOGEE, which measured a velocity dispersion of $\sim$1 km s$^{-1}$, yielding a virialized cluster with some initial substructure \citep{foster2015}. The APOGEE project also observed IC 348 measuring a velocity dispersion of $\sim$0.7 km s$^{-1}$, and implying a supervirial state; consistent that IC 348 is in an advanced phase of gas dispersal \citep{cottaar2015}. \citet{ortizleon2018} used the GAIA DR2 release on IC 348 and NGC 1333 and found velocity dispersions that agree with the APOGEE project. 

The Gaia-ESO survey studied several young regions probing their kinematical structure: \citet{jeffries2014} determined there were two kinematical populations in $\gamma$ Vel B with velocity dispersions of 1.6 and 0.34 km s$^{-1}$, \citet{rigliaco2016} determined a velocity dispersion of 1.14 km s$^{-1}$ in L1688 in the $\rho$ Ophiuchi molecular cloud, and \citet{sacco2017} records a velocity dispersion of 1.1 km s$^{-1}$ in Chamaeleon I. For post gas dispersal clusters, \citet{jeffries2006} determined there are two spatially superimposed components in $\sigma$ Ori with velocity dispersions of 1.1 and 1.3 km s$^{-1}$, and \citet{kuhn2014} studied NGC 6231 in the Sco OB1 association and found it is gravitationally bound. 

In \S2, we describe our observations and data reduction. In \S3, we provide our analysis,
and \S4 contains the results. We discuss implications of the results in \S5 and summarize the paper in \S6.

\section{Hectoschelle Observations and Data Reduction}
\label{sec:observations}
The spectra were obtained on the MMT with Hectoschelle \citep{szentgyorgyi1998}, a fiber-fed, echelle spectrograph with resolving power of $\sim$32,000. The 240 fibers are robotically placed on the maximum possible number of target stars for a given configuration. Five epochs of observations were taken over five years (2009-2013, see Table 1) using the 25$^{th}$ order; these spectra span 150~\AA~centered on H$\alpha$ at 6563~\AA. Every configuration had $\sim$30 fibers placed on the sky to measure the sky emission and the remaining fibers placed on target stars with integration times of 4 x 1800 seconds, except for the final epoch, which was 4 x 2100 seconds. In addition, one set of spectra were obtained with the telescope offset by 5$\arcsec$ from the target stars per configuration to measure the contribution of interstellar lines in the spectra.

The target stars are from the combined visible, infrared, and X-ray study of \citet{allen2012}. They determined membership by the detection of an IR-excess with $Spitzer$ IR using color-color diagrams, by the detection of X-ray emission from enhanced coronal activity using novel and archival $Chandra$ data, and - less reliably - by their coincidence in $V$ vs. $V$ - $I$ diagrams with the isochrone of members identified with the X-ray and $Spitzer$ data. Diskless members with X-ray detections were the highest priority targets of the survey since H$\alpha$ emission dominates the spectrum of disk objects due to accretion. $Spitzer$-identified YSOs with disks, both Class II and transition-disk objects, were given second priority. Third priority was assigned to the potential members identified in the $V$ vs. $V-I$ color magnitude diagram; these would be diskless members without detectable X-ray measurements. The lowest ranked objects were objects with a V-band magnitude of $\sim$11.5 to 14.5 regardless of the potential for membership. These objects were targeted if no more fibers could be placed targeting any of the three higher ranked objects. A total of 561 spectra (over five epochs, see Tables 2-6) were taken toward Cep OB3b of 499 distinct stars, 190 of which were identified as members using the above criteria.

All spectra were reduced following the procedure in \citet{szentgyorgyi2006}, which uses the standard IRAF reduction pipeline for echelle spectra. Cep OB3b has a significant amount of nebular emission stemming from the S 155 HII region nearby. Spectra of the nebulosity taken at offset positions were used to subtract out the contribution from the nebular emission, as described in Prchlik et al. (submitted).

\section{Extraction of Radial Velocities}

We used the $rvsao$ package \citep{kurtz1998} within IRAF to determine the RV of each spectrum. Synthetic spectral templates from \citet{munari2005} were adopted for the cross correlation. The templates had [Fe/H] = $+$0.5 solar using the stellar atmosphere models from \citet{kurucz1993} that covered 2500 - 10500~\AA~with a resolving power of 20,000. A super-solar metallicity was chosen to account for higher metallicity in younger stars compared to the Sun. Each spectrum was cross correlated with eleven templates ranging from 3500~K to 6250~K in steps of 250~K. We also cross correlated with the higher resolution synthetic templates of \citet{coelho2005}. These provided solutions consistent with those using Munari et al., hence we adopted the RV values using the Munari et al. templates because they were closer in resolution to our data. The H$\alpha$ region (6559 $-$ 6566~\AA) was masked out of all spectra regardless of strength to avoid contamination of the extracted RV. We did not account for potential broadening of the lines by stellar rotation. Regions in the spectra with high nebular contamination were also masked out individually by eye to avoid contamination of the measured RV. 

The best matching template \citep[highest R value,][]{tonry1979} was selected yielding the RV and uncertainty used in this analysis. The R value corresponds to the quality of the cross-correlation and scales as R$=\frac{h}{\sqrt{2}\sigma_{a}}$, where $h$ is the height of the cross correlation peak and $\sigma_{a}$ is the estimated uncertainty from the rms of the antisymmetric portion of the correlation function. A RV range of -50 to +50 km s$^{-1}$ was used as an initial restriction. Each cross correlation plot was inspected by eye. Some objects had a strong peak outside of this range. For these objects the range was extended and run again and the highest R value template result was adopted. Inspection by eye also enabled us to identify objects that were potential binaries. The RV values are given in Table 2 - 7; all reported RVs are in terms of the local standard of rest, V$_{lsr}$. 

The fifth epoch, 2013, observed 64 objects for a second time in addition to 13 objects not previously observed. This was useful to find binaries in the sample in addition to testing the consistency of the velocity calibration. 

Figure 1 shows the difference in RV vs the minimum R value of 42 objects that were observed twice and were not initially suspected as binaries based on the shape of their cross correlation plots (see Sec \ref{sec:binaries}). For 19 stars, we find a consistent offset between the 2009 epoch RVs compared to the 2013 measurement. This offset is not present when comparing the 2010 and 2011 RVs to the 2013 RVs for the 23 stars observed in those epochs. 

In order to quantify the offset, we calculated a weighted average of the RV change between epochs with R of six or higher yielding 5.5 km s$^{-1}$ with an error on the mean of 0.19 km s$^{-1}$. Normally a $\sim$5 km s$^{-1}$ difference between epochs would indicate a binary (assuming the uncertainties are less) since the offset is apparent in most of the sources with high R values. That is not the case for these specific objects. In contrast, the objects that were originally observed in 2010 or 2011 and observed again in 2013, a weighted offset was found to be 0.1 km s$^{-1}$ with an error on the mean of 0.1 km s$^{-1}$. This offset is easily within the uncertainties of the data and therefore no offset correction was needed for the 2010 and 2011 epochs. The 5.5 km s$^{-1}$ offset was applied to both 2009 epochs and the uncertainty of the offset was combined in quadrature with the velocity uncertainties of each object in the 2009 data. These corrected values and the combined uncertainties are used in the tables and analysis. 

The RV measurements of all observations are listed in Tables 2 - 6. The origin of this offset remains unclear. The uncertainties obtained from our analysis are consistent with previous results from Hectoschelle \citep{furesz2006,furesz2008,tobin2009,tobin2015,kounkel2017}. 

\subsection{Identifying Binaries}
\label{sec:binaries}
Potential spectroscopic binaries are listed in Table 7. Binaries of similar mass give a double peak correlation plot, which was seen in 15 objects. For mass ratios less than unity the primary component will dominate the spectrum and thus the RV will measure the motion of the primary. The secondary component may contaminate the spectrum enough to add an asymmetry. A total of 39 objects with significant asymmetries in their cross correlation curves were also marked as potential binaries. If a source was observed twice, we compared the RVs to search for additional binaries. A shift in RV could be noticeable for short to moderate period binaries i.e., orbital periods ranging from weeks up to $\sim$10 years. For objects with no companion, the RV should remain the same within the uncertainties. Of the 64 objects observed for a second time 34 disagree at the $> 1$~$\sigma$ level and 21 disagree at the $> 2$~$\sigma$ level. It is expected that at least 20 objects would differ at the 1$\sigma$ level even if none are binaries, the larger number of objects in disagreement suggests a short-period binary frequency of $\sim$20\% in our sample. We identify all of the $> 1\sigma$ objects as potential binaries.

\begin{figure*}[ht]
\centering
\plotone{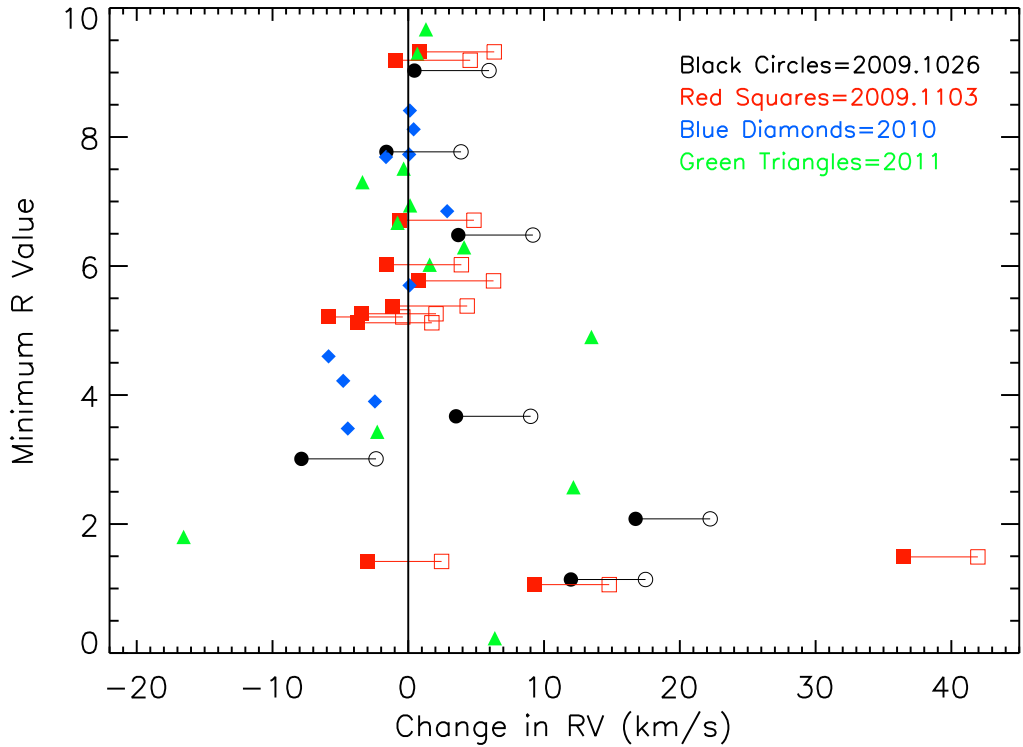}
\caption{Minimum R value vs. change in RV for objects observed in two epochs. The x-axis is the difference in RV between the two epochs. This excludes objects that were flagged as binaries in the initial cross correlation. The observations from 2009.1026/1103 (black circles/red squares, respectively) have a systematic non-zero change in RV compared to the 2010 (blue diamonds) and 2011 (green triangles) epochs at high R. The RV results from both 2009 epochs were shifted by an offset of 5.5 km s$^{-1}$; the open symbols show the change in the RV, the filled symbols connected to these by a line show the change after the offset was applied.} 
\label{fig:matches}
\end{figure*}

\subsection{The RV Analysis Sample}
\label{sec:rvanalysis}

To reliably measure the velocity structure and dispersion of the Cep OB3b sub-clusters, we make three cuts to the RV data to ensure the reliability of the velocities. The first cut eliminates potential binaries as indicated in \S \ref{sec:binaries} (see Table 7). This cut reduces the sample from 190 to 109. The second cut utilizes the R value from $rvsao$, which is a measure of the S/N. We reject objects that have a R $<$ 5 (Figure 2). The value adopted for this cut comes from an analysis of the threshold R value vs the cumulative velocity dispersion, which examined the trade off between sample sizes and the uncertainties in the RVs. In this analysis, the velocity dispersion was calculated for all the non-binary members at or below the R value of the bin (Figure 2). We adopted the threshold for R, which gave the minimum velocity dispersion for a sample size exceeding 30 objects; this minimum occurred at R$\sim$5. Increasing the threshold reduces the number of objects leading to fluctuations from small numbers statistics. A lower threshold includes objects with poorly constrained RV values with inherently large uncertainties associated. This cut reduces the sample from 109 to 62. The third cut is 3-$\sigma$-clipping from the average velocity of the 62 remaining members the cluster to reduce the contamination by nonmembers. The cut leaves 57 objects remaining.

\begin{figure*}[ht]
\centering
\plotone{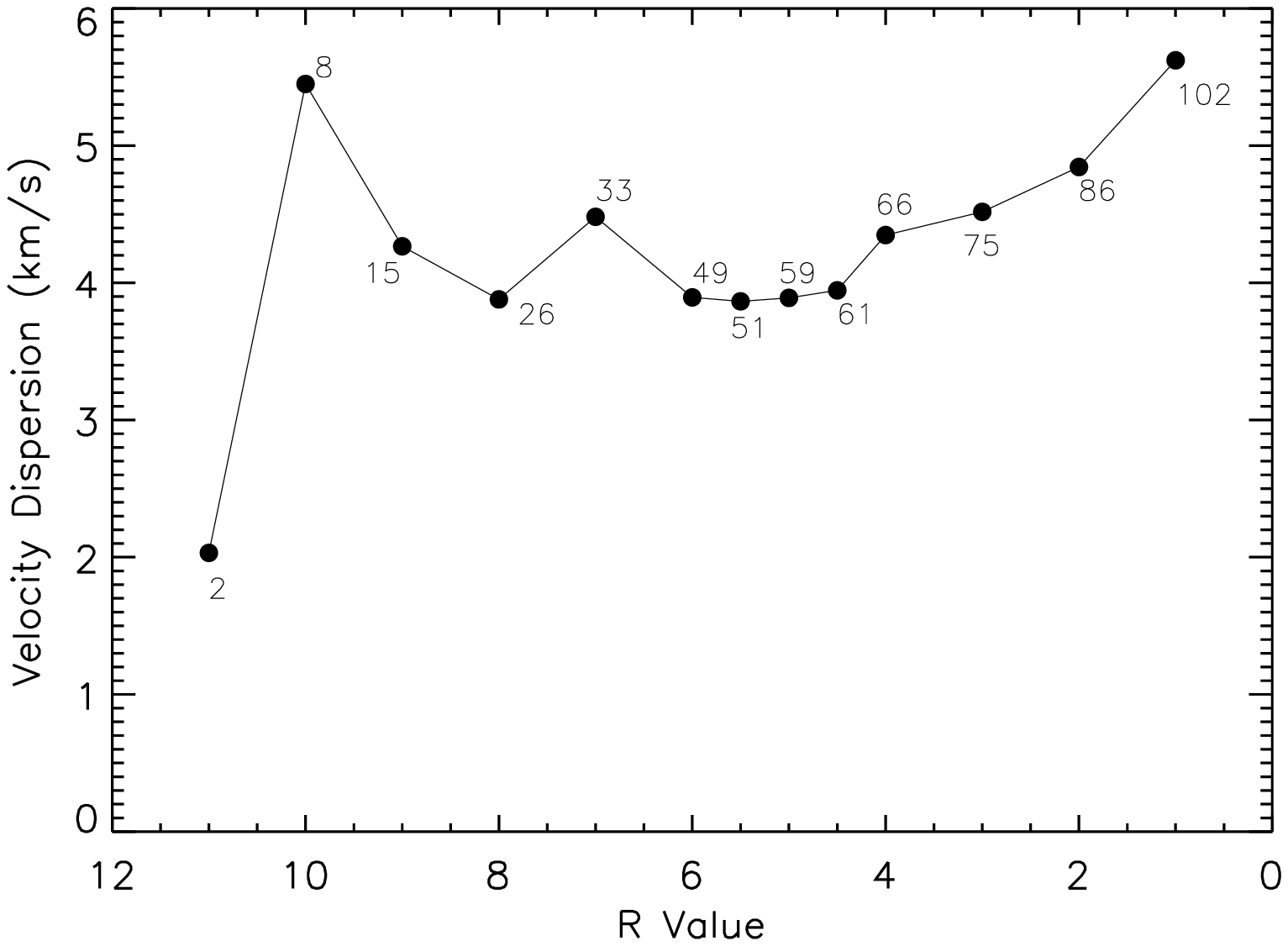}
\caption{The velocity dispersion vs. minimum R value of the stars with youth indicators, which are not identified as binaries. The number of objects included in the calculation is next to each R value point. The velocity dispersion has a broad minimum around R$\sim$5, which is adopted as the R cutoff in \S\ref{sec:rvanalysis}.}
\label{fig:rvalue}
\end{figure*}

\section{Results}

In the following section, we focus on the properties of the two sub-clusters. The rationale for this approach is based on two observations. First, while the overall cluster is highly elongated and hierarchical in structure, each of the two sub-clusters has a single, centrally condensed core and are much closer to circular symmetry \citep[Figure 4 in][]{allen2012}. This suggests that these are two distinct components within the larger cluster that can be studied independently. Furthermore, each sub-cluster is associated with a distinct molecular clump within the larger parental clouds \citep[Figure 4 in][]{allen2012}. After using RVs to assess membership, we first characterize the structure of the two sub-clusters, fitting them to the empirical \citet{king1962} model to constrain their three dimensional structure. We then analyze the RV structure and velocity dispersion of each sub-cluster. Finally, we use Gaia DR2 to determine distance of Cep OB3b and bulk proper motions of the sub-clusters. From these results, we can also estimate the gravitational potential energy of the sub-clusters. These properties will inform a discussion of the dynamical status and fate of the sub-clusters in subsequent sections. 

\begin{figure*}[ht]
\centering
\plotone{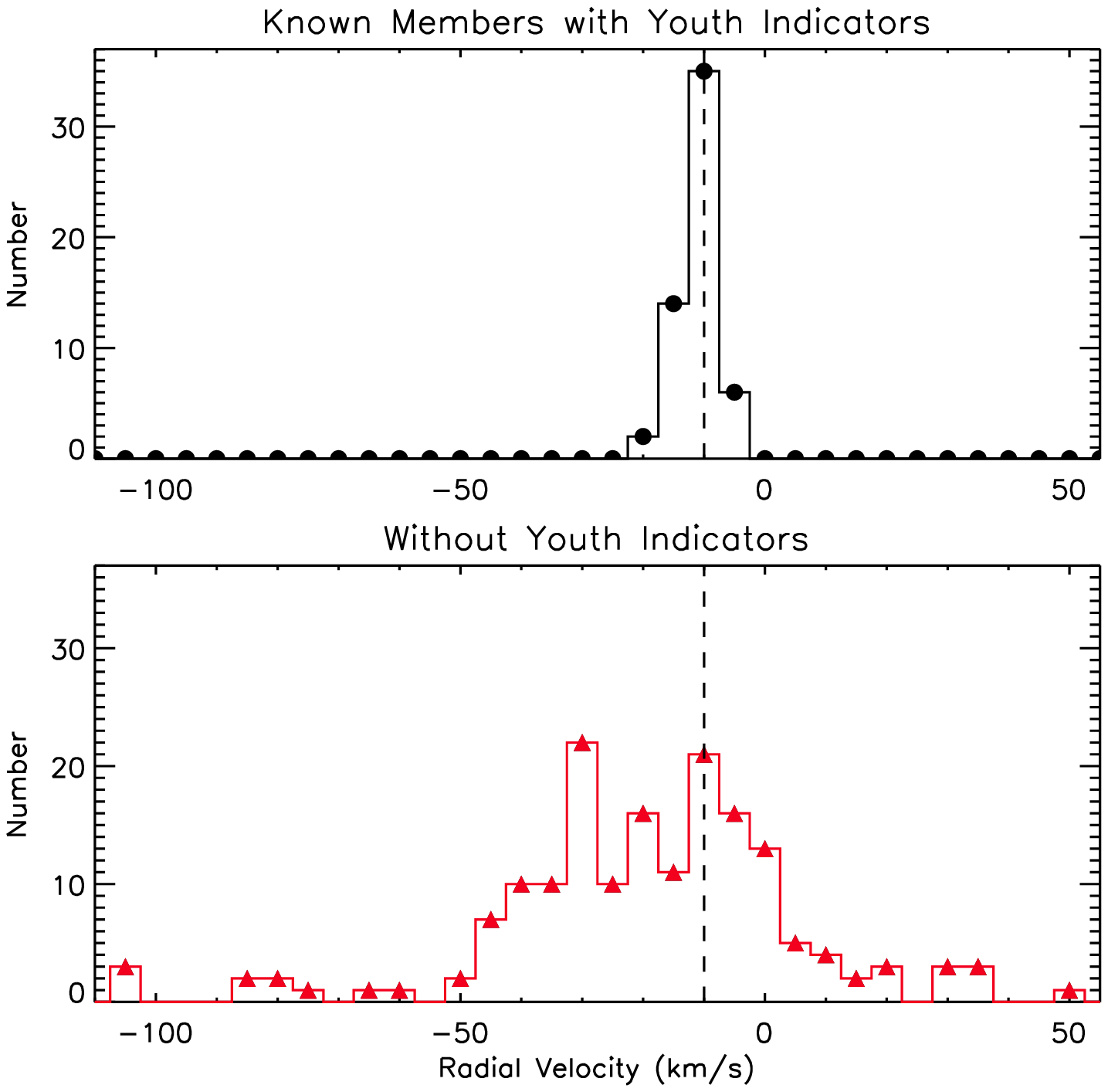}
\caption{Radial velocity distribution of objects with youth indicators (top) and objects lacking youth indicators (bottom). The molecular cloud peaks at -10 km s$^{-1}$, which agrees with the LSR of the molecular cloud. More local clouds like Cep OB3b are found closer to 0 km s$^{-1}$. Objects with highly negative RVs are likely background stars, with objects in the Perseus arm peaking at -30 km s$^{-1}$.}
\label{fig:histo}
\end{figure*} 

\subsection{RV Analysis of Membership}

Due to the low extinction through the cluster, Hectoschelle can detect both foreground and background objects in the field in addition to members of the cluster. We therefore break up the sample into two categories: objects with youth indicators and objects without youth indicators, most of which are likely contaminants. Youth indicators are the presence of an infrared excess, implying the presence of a dusty disk or envelope (i.e., class II objects, transition disk objects, or protostars), or detectable X-ray emission due to an active coronae. The objects with X-ray emission but no IR-excess are diskless pre-main sequence (pre-ms) stars (class III objects). The velocities of the stars with youth indicators show a clear peak at an RV of -10~km~s$^{-1}$ bin with a narrow RV distribution (Figure 3).

Objects without youth indicators may also be class III objects; however, such objects are indistinguishable from background giants or foreground dwarf objects on the basis of photometry alone. Figure 3 shows histograms for objects with youth indicators and objects without youth indicators. The objects with youth indicators show a clear peak at -10 km s$^{-1}$. In contrast, the histogram of the objects without youth indicators (169)  shows a broad distribution of RVs ranging from -107 to 50 km s$^{-1}$. There is a peak at -30 km s$^{-1}$, which is consistent with the RVs of stars in the Perseus arm behind Cep OB3b. At the velocity of the cluster, -10 km s$^{-1}$, the distribution of the objects without youth indicators is relatively flat compared to the youth indicators; consistent with objects dominated by field stars. Thus, we see no clear evidence for a large number of missed class III objects lacking X-ray detection in the RV distribution, and we do not use RV as an additional criteria for membership.  
 
\citet{heyer1996} imaged $^{13}$CO and $^{12}$CO $J$ $=$ 1$-$0 emission from visible HII regions, which included Cep OB3b. The $^{12}$CO gas toward the cluster was measured at -12.7, -11.4, -10.1, -8.8, and -7.5 km s$^{-1}$, which is consistent with the motions of the members. We do not see a significant offset between the stellar motions and the gas motions, further confirmation of their association with the cloud.

\subsection{Sub-clusters Sizes, Structures and, Densities}
\label{sec:king}

To characterize the structures of the sub-clusters, we apply an analytic model developed by \citet{king1962} for globular clusters. The models are parameterized by a core radius, an outer tidal radius, and a central peak density. Although this model is not appropriate for the elongated and irregular structure of more deeply embedded clusters such as the ONC \citep{kuhn2014,megeath2016}, the more circular symmetry of the two Cep OB3b sub-clusters motivates the use of this model. We use the $Chandra$ and $Spitzer$ data for the King model fits. We adopt the center positions of the sub-clusters given by the method of \citet{gutermuth2009}. The algorithm isolates the two most numerous YSO overdensities that lack further substructure using the catalog and field of view limits of \citet{allen2012}. We compute the azimuthally average radial surface density profile in 0.3 pc bins and fit them with King models using the IDL implementation of {\it mpfit}, specifically the mpcurvefit.pro module \citep{markwardt2009}.

\begin{figure*}[ht]
\centering
\plotone{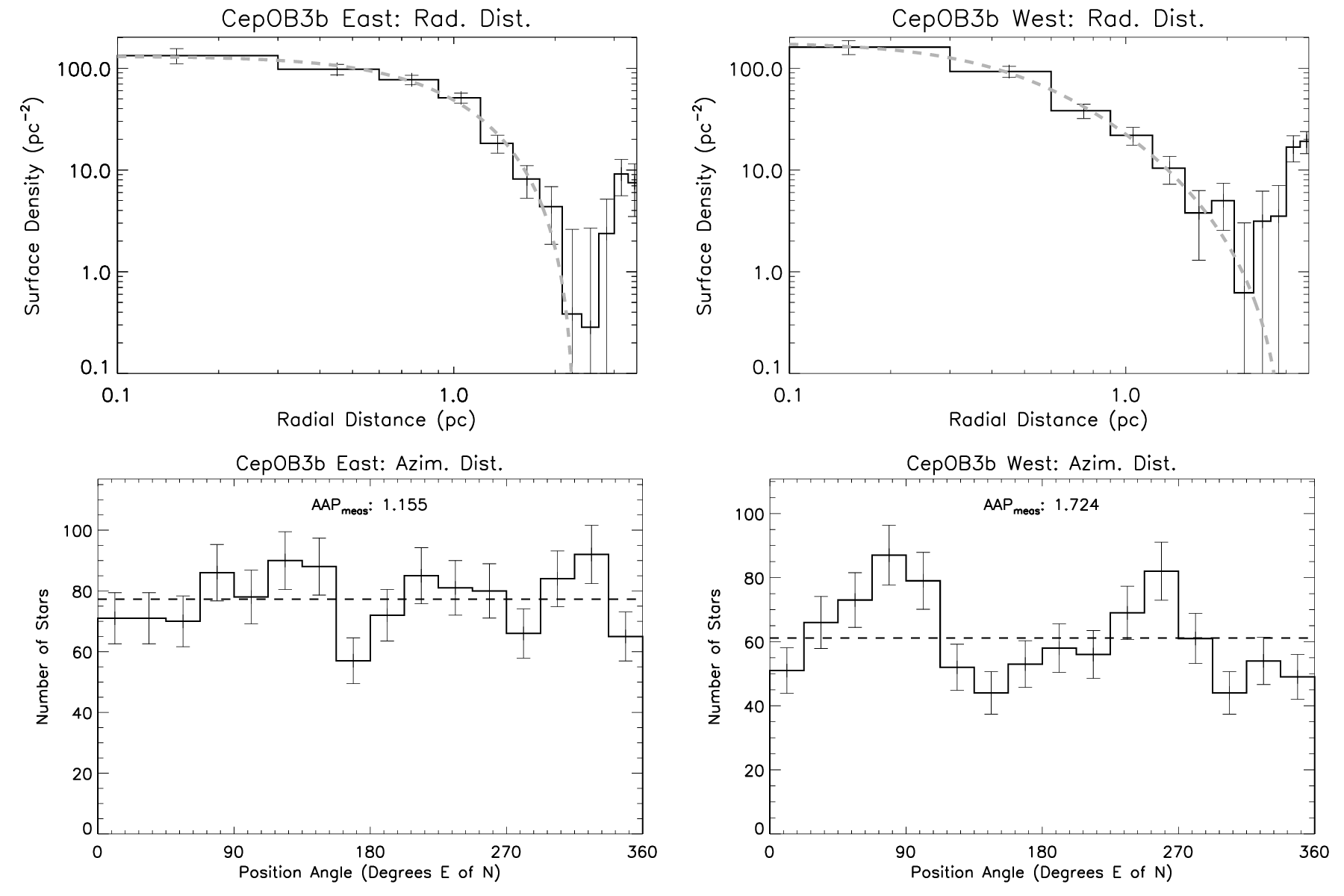}
\caption{The best fit 1962 King models. The top left and right plots are surface density vs. radial distance for the east sub-cluster and west sub-clusters, respectively. The bottom left and right plots are the number of stars vs. position angle for the east and west sub-clusters, respectively. The deviation of these from circular symmetry, which is given by the dashed line, determines the value of the azimuthal asymmetry parameter (AAP) \citep{gutermuth2005}.}
\label{fig:kingnew}
\end{figure*} 

Since the larger annuli extended in part past the $Chandra$ field of view, we must correct for the missing $Chandra$ objects. We do this by measuring the density of the X-ray objects for the section of the annulus within the $Chandra$ field of view. In addition, we must correct for the stars which did not have infrared excesses and did not have bright enough X-ray emissions to be detected by $Chandra$. \citet{allen2012} determined disk fractions in each sub-cluster using visible color-magnitude diagrams and the X-ray data \citep[see Table 5][]{allen2012}. We adopt the upper and lower values of the disk fractions for each sub-cluster to correct for missing objects. We do this by determining a correction factor for each sub-cluster, given by the equation

\begin{equation}
	\eta = \frac{1}{f_{disk}}\frac{N_{IR}}{N_{IR}+N_{X-ray}} ,
\end{equation}
\noindent where $N_{IR}$ is the number of IR$-$excess objects with the boundary of the given sub-clusters, $N_{X-ray}$ is the number of $Chandra$ objects, and $f_{disk}$ is the disk fraction. This assumes that $N_{IR}/f_{disk}$ gives the total number of objects. Finally, we take the total number of stars from the $Chandra$ and $Spitzer$ data sets in each sub-cluster and subtract off a 15.5 pc$^{-2}$ baseline density. This was done to 1) exclude the larger scale halo of YSOs that are likely part of the larger Cep OB3b cluster and 2) to keep the two sub-clusters separate. After the 15.5 pc$^{-2}$ baseline density is subtracted, the measured densities are multiplied by the correction factor to determine the corrected density of stars in each annulus. This yields 809 and 664 stars for disk fractions of 0.32 and 0.39, respectively, in the east and 501 and 416 stars for disk fractions of 0.44 and 0.53, respectively, in the west. The number of objects determined here are used in the potential energy calculations. 

The fits give different results for the two sub-clusters (Figure 4, Table 8). The core radii are 1.36$\pm$0.30 pc and 0.52$\pm$0.11 pc for the east and west, respectively. The tidal radii are 2.32$\pm$0.19 pc and 3.1$\pm$1.0 pc for the east and west, respectively. We note that the tidal radius usually refers to the radius at which the tidal field of the galaxy dominates over the gravity of the cluster, separating stars out of the cluster. Here the tidal radius defines the edge of the sub-clusters in the King model and we will refer to it as the sub-cluster radius throughout the remainder of the paper.

In the above analysis, we have assumed the two sub-clusters are circularly symmetric. To test how closely the clusters follow this assumption, we compute the Azimuthal Asymmetry Parameter, or AAP, from \citet{gutermuth2005} for the two sub-clusters. The value of the AAP measures the deviation of the azimuthal distribution of stars from circular symmetry. We find values of 1.155 and 1.724 for the east and west, respectively. This shows that the surface density distribution of stars in the west deviate significantly from circular symmetry and that the sub-cluster is significantly elongated. In comparison, the east appears circularly symmetric with a relatively smooth distribution of stars. 

The stellar peak densities are 521 and 428 stars pc$^{-2}$ in the east and 342 and 284 stars pc$^{-2}$ in the west, depending on disk fraction. The average stellar surface densities within the core radii are 240 and 197 stars pc$^{-2}$ in the east and 220 and 182 stars pc$^{-2}$ in the west, depending on disk fraction. From these values, the east is significantly richer in members than the west and is circularly symmetric showing a larger core radius. The west is characterized by elongation, a core radius that is less than half of that for the east, and has fewer members. From the King model fits the half-mass radius is 0.83 pc and 0.67 pc for the east and west, respectively. 

\subsection{The Velocity Structure of the Cep OB3b Cluster}
\label{sec:velocitystructure}

The position-velocity (PV) diagrams of the Cep OB3b cluster are shown in Figures 5 and 6. These figures indicate the lack of an apparent velocity gradient in declination or right ascension for both sub-clusters and the entire region itself. The RVs in the west have a relatively flat distribution around the average RV. On the other hand, the east has a much wider spread of RVs but still lacks a RV gradient in right ascension or declination.

The average RV of the two sub-clusters are very similar, -12.09 km s$^{-1}$ with a standard error of the mean as 0.563 km s$^{-1}$ and -10.86 km s$^{-1}$ with a standard error of the mean as 0.538 km s$^{-1}$ for the east and west, respectively. There are within 2$\sigma$ of each other. The overall RV average is -11.69 km s$^{-1}$ with a standard error of the mean as 0.423 km s$^{-1}$. 
\begin{figure*}[ht]
\centering
\plotone{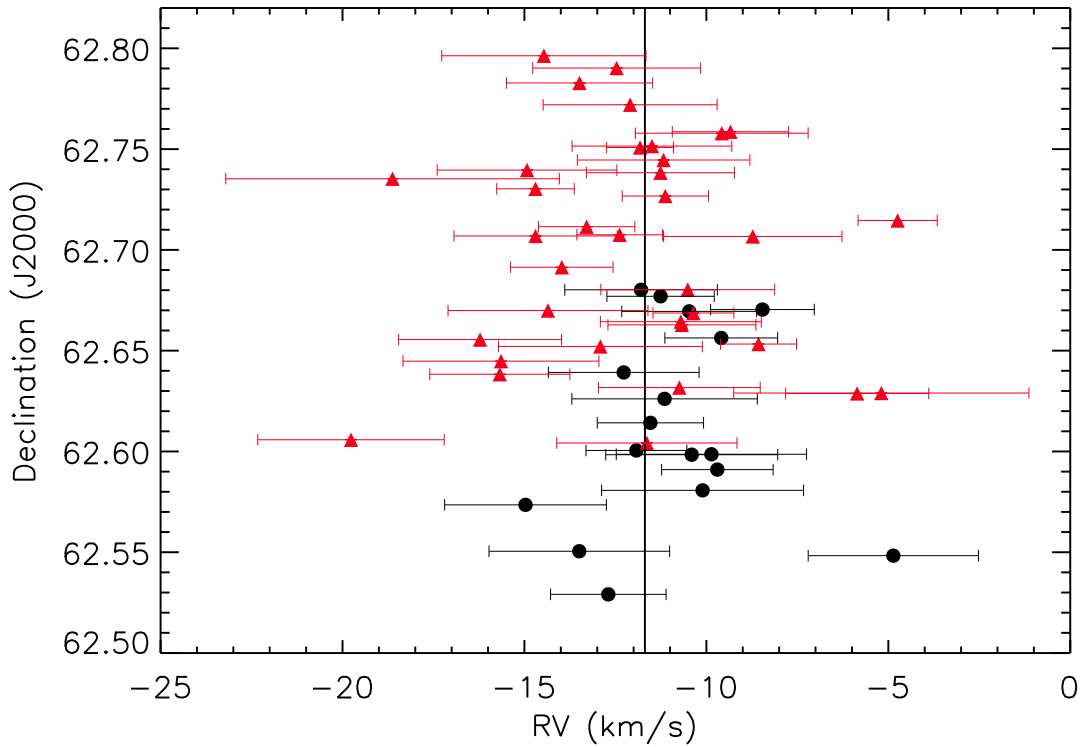}
\caption{Declination~vs.~RV for stars with youth indicators. The red triangles are objects in the east sub-cluster. The black circles are objects  in the west sub-cluster.}
\label{fig:rvdecfinal}
\end{figure*}

\begin{figure*}[ht]
\centering
\plotone{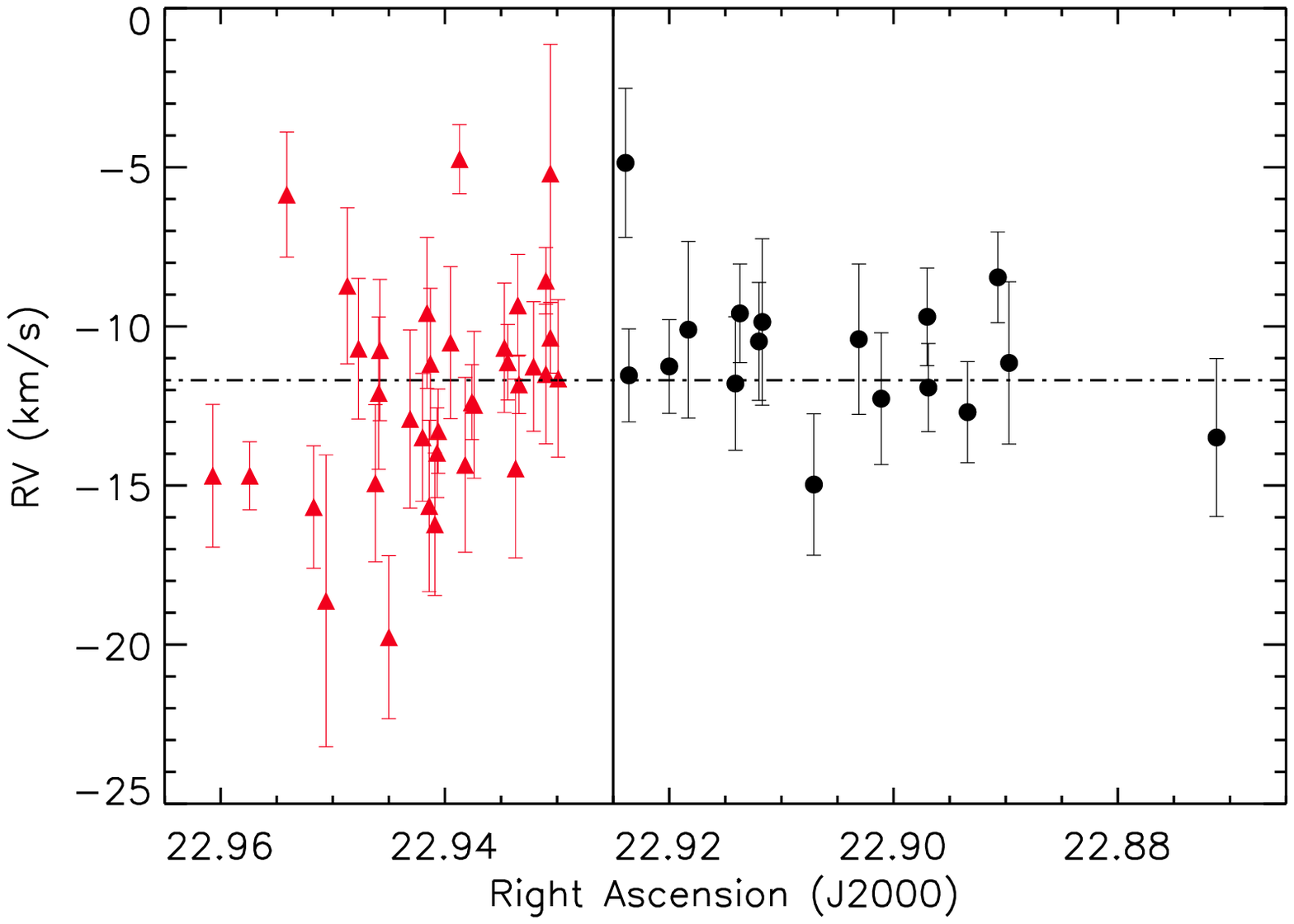}
\caption{Radial velocity vs. RA for objects with youth indicators. The symbols and colors are the same as in Figure \ref{fig:rvdecfinal}.} 
\label{fig:rvfinal}
\end{figure*}

\subsection{The Velocity Dispersion of the Sub-Clusters}

To constrain the velocity dispersions of the sub-clusters, we implement a Bayesian parameter estimation for the velocity dispersion in each sub-cluster. We determine the likelihood function using a Monte-Carlo comparison to our measured velocity dispersions for the 35 and 17 stars of the eastern and western sub-clusters. We start by adopting 1D gaussian velocity distributions centered on the RV of each sub-cluster; the distributions of the sub-clusters have independent velocity dispersions, $\sigma$. To calculate the velocity dispersion, 35 RVs and 17 RVs are randomly drawn from the Gaussian distribution of the east and west, respectively. We iterate this process, changing the width $\sigma$ of the Gaussian distribution. For each sub-cluster, the $\sigma$ starts at 0.05 km s$^{-1}$ and increases in steps of 0.05 km s$^{-1}$ until $\sigma$ = 10 km s$^{-1}$ is reached. The RVs were drawn 10,000 times for each value of $\sigma$.

To account for unresolved binaries, we adopt the approach of \citet{cottaar2012} to add the effect of orbital motions. These would be cases without a double-lined spectrum where the velocities that we measure are those of the more luminous primary stars. We adopt three mass-based (1~M$_{\odot}$, 0.75 M$_{\odot}$, 0.5 M$_{\odot}$) log-normal velocity distributions. The solar-type distribution of absolute velocities has a log-normal width of 0.84 log$_{10}$ km s$^{-1}$ and a mean of 0.08 log$_{10}$ km s$^{-1}$ \citep{cottaar2012}. The 0.75 M$_{\odot}$ (0.62 log$_{10}$ km s$^{-1}$ width and 0.35 log$_{10}$ km s$^{-1}$ mean) and 0.5 M$_{\odot}$ (0.15 log$_{10}$ km s$^{-1}$ width and 0.52 log$_{10}$ km s$^{-1}$ mean) distributions are scaled from the solar-type distribution by implementing equation 3 in \citet{cottaar2012} and using the orbital period distributions from Table 1 in \citet{duchene2013}. Using the spectral types from Allen et al. (in prep), we estimate the masses of the 35 (17) members and assign them to the appropriate mass-based binary velocity distribution. A total of 13 objects fall into the solar-type bin, 15 fall into the 0.75 M$_{\odot}$ bin, and 24 fall into the 0.5 M$_{\odot}$ bin.

We test three binary fractions: 0, 0.5, and 1. Stars are randomly assigned a binary motion from the appropriate mass-based Gaussian distribution. The typical FWHM of the cross correlation plots are $\sim$40 km s$^{-1}$. We throw out binary motions that were greater than this because we are able to detect them in the initial RV extraction. Finally, to take into account the uncertainties in the RV measurements, we add velocities drawn from randomly sampling Gaussian distributions with $\sigma$ equal to the uncertainties of the measured velocities.

We use the fraction of times the 10,000 simulated velocity distributions are within 0.1 km s$^{-1}$ of the actual velocity distribution to create the posterior probability density function (PDF) of the velocity dispersions. The PDFs of the velocity dispersions for the east and west are shown in Figures 8 \& 9, respectively. The expectation value for the two sub-clusters for a binary fraction of 0.5 with $\pm$1~$\sigma$ confidence limits are $1.91^{+0.50}_{-0.42}$ km s$^{-1}$ for the east and $1.10^{+0.30}_{-0.44}$ km s$^{-1}$ for the west. The peak values for these distributions are 2.2 km s$^{-1}$ and 0.5 km s$^{-1}$ for the east and west, respectively.

We explore the influence of the binary fractions on the velocity dispersion of the eastern sub-cluster in Figure 8. The binary fraction changes the resulting velocity dispersion for the east. The velocity dispersion is centered at 2.8 km s$^{-1}$ for a binary fraction of zero. As the binary fraction increases the distribution widens and flattens moving closer to 0 km s$^{-1}$. This demonstrates that accounting for unresolved binaries is an important step to probing an accurate kinematical survey of young clusters. The western sub-cluster probability distribution (Figure 9) peaks at 1.5 km s$^{-1}$ for zero binaries and moves closer to zero for binary fractions of 0.5 and 1.0 flattening and widening as in the east (Figure 8). 

\citet{jeffries2014} uses a maximum likelihood technique to account for unresolved binaries in Gamma Vel, which is adopted from \citet{cottaar2012}. Applying this approach to the Cep OB3b data, with the binary fraction set to 0.5 and the masses at 0.5 M$_{\odot}$ results in an intrinsic Gaussian dispersion of 1.62$\pm$0.75 km s$^{-1}$ and 0.78$\pm$0.60 km s$^{-1}$ for the east and west, respectively. If the binary fraction is set to 0 then the dispersions increase to 2.54$\pm$0.44 km s$^{-1}$ and 0.8$\pm$0.61 km s$^{-1}$ for the east and west, respectively. The results of the velocity dispersions for the different binary fractions agree at the 1$\sigma$ level between our bayesian analysis and the maximum likelihood technique. 

\begin{figure*}[ht]
\centering
\plotone{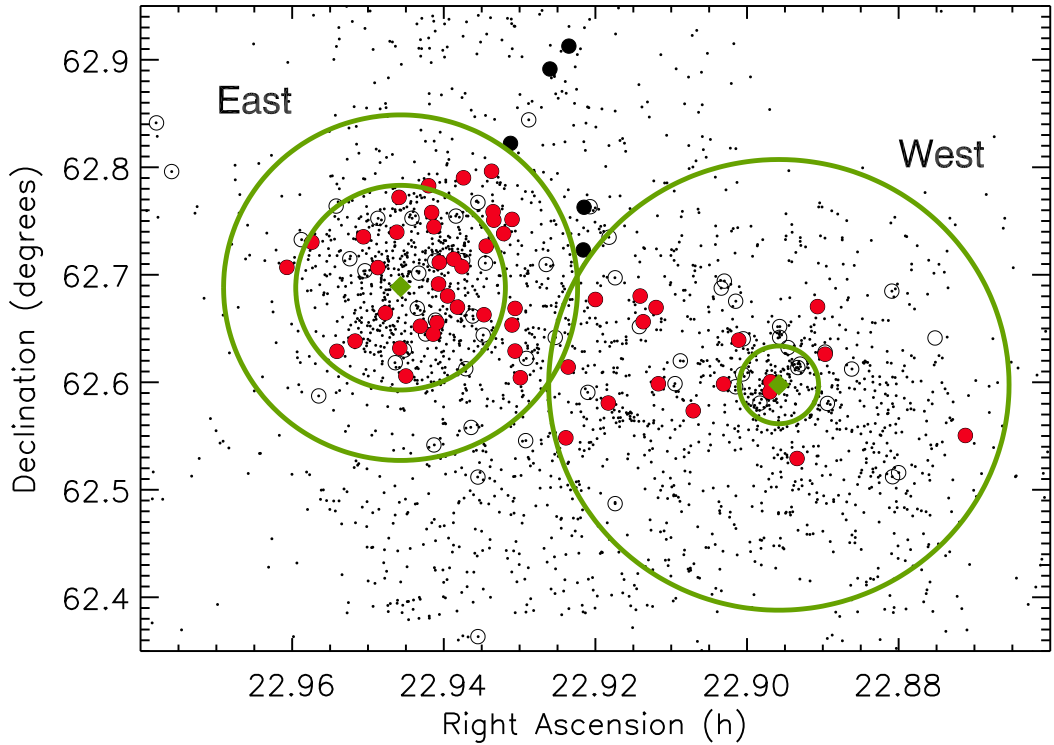}
\caption{The black points are all the members of Cep OB3b \citep{allen2012} and the open circles have Hectoschelle spectra. The filled black circles are members with Hectoschelle spectra that survived the cuts described in \S \ref{sec:observations}, but fall outside the sub-cluster radii. The filled red circles are members with Hectoschelle spectra within the sub-cluster radii and are used in the kinematic analysis. The green diamonds are the centers of the respective sub-clusters and the green circles are the core and sub-cluster radii for each sub-cluster.}
\label{fig:tidalmembers}
\end{figure*}

\subsection{The Total Energy of the Eastern and Western Sub-Clusters}
\label{sec:total_energy}

We use equation 27 in \citet{king1962} and combine the parameter fits of the core radius, sub-cluster radius, and the number of stars from \S \ref{sec:king} to determine the potential energy of the sub-clusters. The number of stars and the sub-cluster radius are kept constant and only the core radius was allowed to vary because it has the largest impact on the potential energy. The value of A, the surface density of stars at the center of the cluster, is adjusted to keep the number of stars fixed to the total number of stars estimated in \S \ref{sec:king}; the uncertainty in this number is taken into account by repeating the calculation with the two different disk fractions. We ran the calculation 10,000 times to create a PDF for the potential energies of each sub-cluster. 

We adopt the PDFs of the sub-cluster velocity dispersions to derive kinetic energies of the sub-clusters 10,000 times. The kinetic energy PDFs assume a symmetric three dimensional velocity dispersion with the $\sigma$ for the two directions in the plane of the sky equal to that in the radial direction. Thus, the total kinetic energy is given by 3/2 $\sigma^{2}$. Note that the mass of stars (0.5 $M_{\odot}$$\times$ $N_{members}$) appears linearly in the denominator of T/$\mid$U$\mid$ because the mass in the kinetic energy cancels a mass in the potential energy.  

For each of the 10,000 iterations, we combine the kinetic and potential energy PDFs for each sub-cluster to determine a final PDF of 10,000 points, i.e., log(T/$\mid$U$\mid$). We plot the quartiles of this distribution in Figure 10. The change in the resulting value of log(T/$\mid$U$\mid$) between the two disk fractions for both sub-clusters is insignificant and therefore we show one disk fraction for each sub-cluster in Figures 10 \& 11. Adopting a binary fraction of 0.5, the east log(T/$\mid$U$\mid$) has a mean value at 0.3 and a value at the peak of the distribution of 0.6 (Figure 11). This implies that the eastern sub-cluster is unbound and in a state of expansion. A binary fraction of 0 results in log(T/$\mid$U$\mid$) $>$ 0, also implying an unbound, expanding state. We note that unbound simply means that log(T/$\mid$U$\mid$) $>$ 0; it is possible that parts of the sub-cluster may be bound as we will discuss later. A binary fraction of 1 has a mean value in a subvirial, bound state with a probability of 61\% of being unbound.

For the west sub-cluster, adopting a binary fraction of 0.5, results in a mean of log(T/$\mid$U$\mid$)$\sim$-0.16 with a large range of outcomes and a peak value at 0.3 (Figure 11). The mean value falls into an approximate virial state with a 55\% chance of an unbound state. A binary fraction of 0 results in a subvirial state with 79\% probability of being unbound. For a binary fraction of 1, the west is subvirial with roughly equal chances of being bound or unbound. From these results, the west appears to have a velocity dispersion very close to zero, depending on the binary fraction. 

It is important to recall that not all of the gas mass has been expelled from Cep OB3b even though it is currently in the gas dispersal phase. Gas remains in both sub-clusters \citep[see][Figure 4]{allen2012}. We calculate the amount of gas mass in the sub-clusters based on the 2MASS extinction map. \citet{allen2014} calculated there was roughly 1 A$_{v}$ of foreground extinction along the line of sight to Cep OB3b. We integrate the amount of extinction greater than 2 mag inside each sub-cluster radius and adopt a distance of 819 pc from Gaia DR2 as explained below in \S \ref{sec:propermotion}. This results in $\sim$136 M$_{\odot}$ and $\sim$697 M$_{\odot}$ of gas mass in the east and west, respectively. The gas measured in the east is concentrated near the edge of the sub-cluster and is not centrally located. In contrast, the gas mass in the west is concentrated inside the core radius. The west's mass is dominated by the gas, up to 77\%, and because it is centrally concentrated it is necessary to include the gas mass in the potential energy. Accounting for the gas mass, the log(T/$\mid$U$\mid$) values drop by $\sim$0.4 resulting in a subvirial, bound state, which we adopt as the more accurate kinematical result. 

\begin{figure*}[ht]
\centering
\plotone{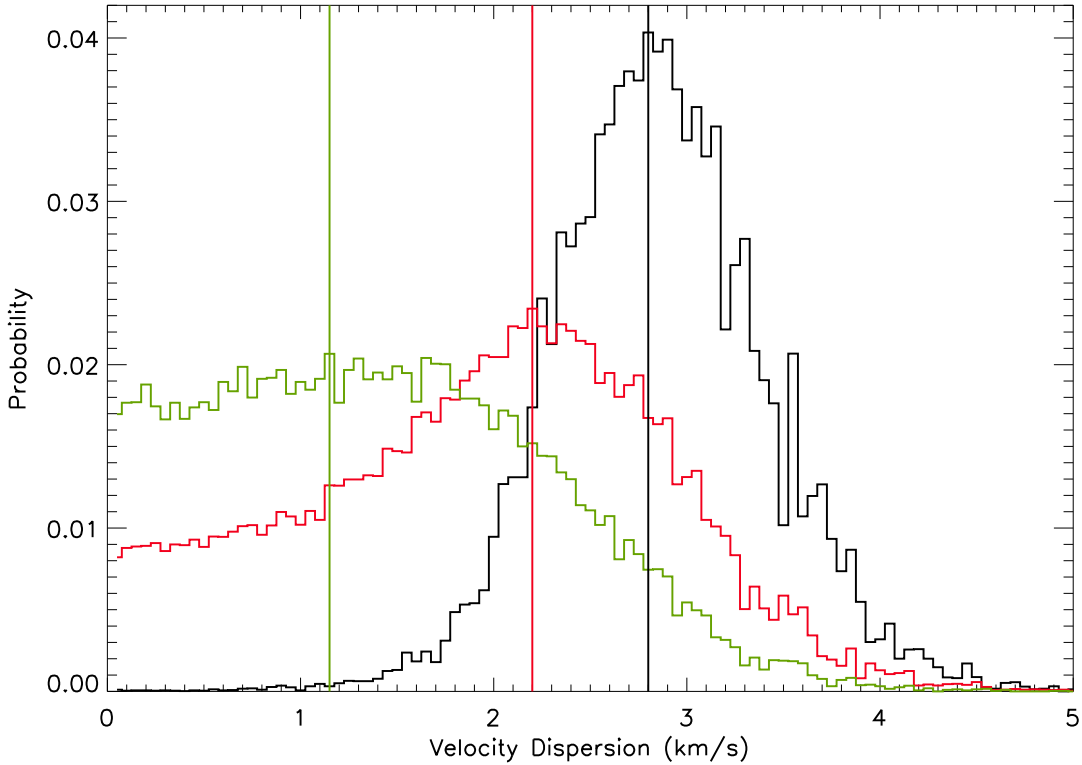}
\caption{The PDF of the velocity distribution in the east sub-cluster. The black histogram is the PDF for a binary fraction of 0 with a peak velocity dispersion value of 2.8 km s$^{-1}$, the red for a binary fraction of 0.5 with a peak velocity dispersion value of 2.2 km s$^{-1}$, and the green for a binary fraction of 1.0 with a peak velocity dispersion value of 1.15 km s$^{-1}$. The vertical lines represent the peak value of the distribution. Changing the binary fraction has a significant effect on the PDF of the velocity dispersion.}
\label{fig:eastveldisp}
\end{figure*}

\begin{figure*}[ht]
\centering
\plotone{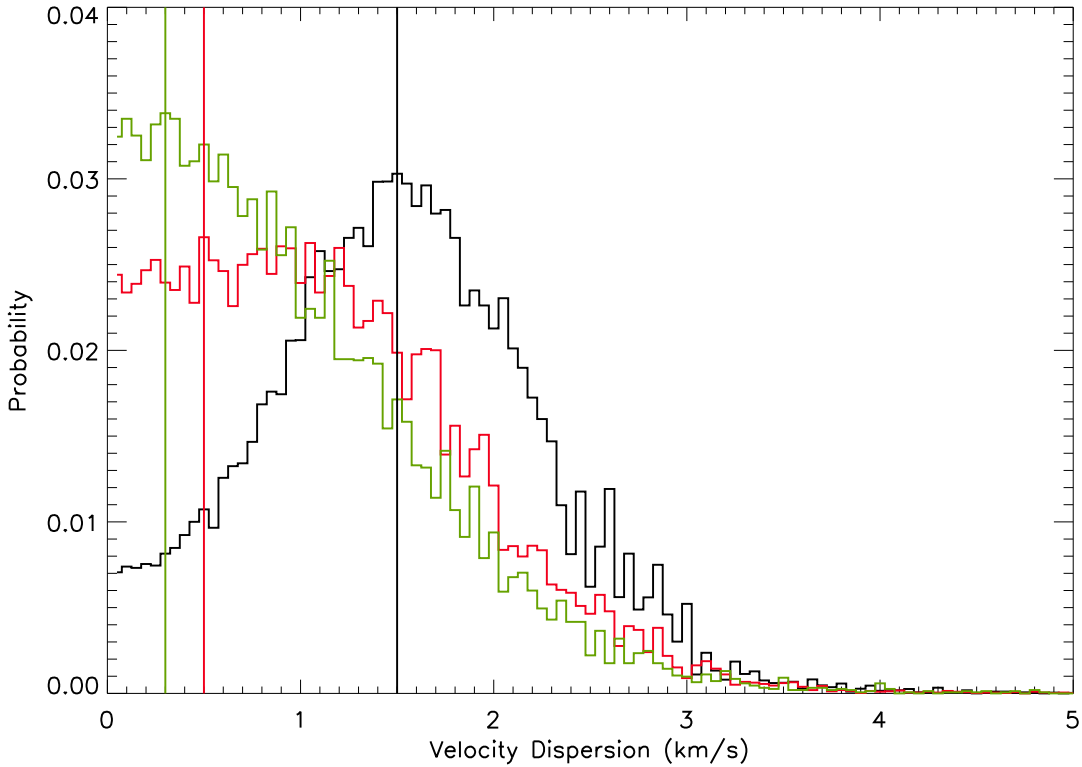}
\caption{The PDF of the velocity distribution in the west sub-cluster. The black histogram is the PDF for a binary fraction of 0 with a peak velocity dispersion value of 1.5 km s$^{-1}$, the red for a binary fraction of 0.5 with a peak velocity dispersion value of 0.5 km s$^{-1}$, and the green for a binary fraction of 1.0 with a peak velocity dispersion value of 0.3 km s$^{-1}$. The vertical lines represent the peak value of the distribution. Changing the binary fraction has little effect on the resulting velocity dispersion.}
\label{fig:westveldisp}
\end{figure*}

\subsection{The Distance to Cep OB3b and the Proper Motions of the Sub-Clusters} 
\label{sec:propermotion}

We cross matched the Gaia DR2 catalog \citep{marrese2018} with CepOB3b members with youth indicators \citep{allen2012} using their 2MASS IDs. We calculate the average proper motion vector components in RA ($\mu$$_{\alpha}*$) and Dec ($\mu$$_{\delta}$), where $\mu$$_{\alpha}*$ is $\mu_{\alpha}\times\cos\left(\delta\right)$. We use a RA of 22:54:48 to split the stars into the two sub-clusters. We perform a weighted fit of a Gaussian to the distributions of proper motions. We included proper motions between -10 and 10 mas yr$^{-1}$ in bins of 0.2 mas yr$^{-1}$. The weight in each bin is given by the number of stars. The uncertainty in the mean proper motion is given by the Gaussian width normalized by the square root of the number of stars in each sub-cluster (696 and 370 for the east and west, respectively). The east has proper motion components of ($\mu$$_{\alpha}*$, $\mu$$_{\delta}$) equal to (-0.59$\pm$0.02, 2.32$\pm$0.02) mas yr$^{-1}$, while the west sub-cluster has proper motion components of (-1.25$\pm$0.02, -2.78$\pm$0.02) mas yr$^{-1}$ (Figure 12). The total cluster has proper motion components of (0.69$\pm$0.02, -2.44$\pm$0.02) mas yr$^{-1}$. These values agree at the 1$\sigma$ level with \citet{kuhn2018}. Figure 13 shows the direction of the proper motions of each sub-cluster with the average proper motion of the total cluster removed (also see Table 9). 

To determine a distance to Cep OB3b, we used the cluster members with less than 20\% uncertainty in their parallax measurements and nearby companions in the 2MASS point source catalog; the resulting histogram of parallaxes show a clear peak (Figure 12). The parallax value correction of 0.029 mas yr$^{-1}$ was used \citep{lindegren2018}; a distance of 819$\pm$16 pc is found after correcting for the zero point offset of the Gaia parallaxes. The recommended DR2 filtering to remove objects with high excess of astrometric noise was also applied to our sample. Previous measurements have been made for the distance to Cep OB3b: 580$\pm$60 pc using an age and distance ladder \citep{littlefair2010}; 700 pc from a maser parallax-determined distance for Cepheus A \citep{moscadelli2009}; 725 pc from near-IR color magnitude diagrams \citep{sargent1977,getman2009}; 850 pc based on $V$ vs. $V-I$ color-magnitude diagram \citep{mayne2007}. The Gaia distance resolves the inconsistency in the distance estimates and places the cluster at the upper end of the range of previous estimates.

\begin{figure*}[ht]
\centering
\plotone{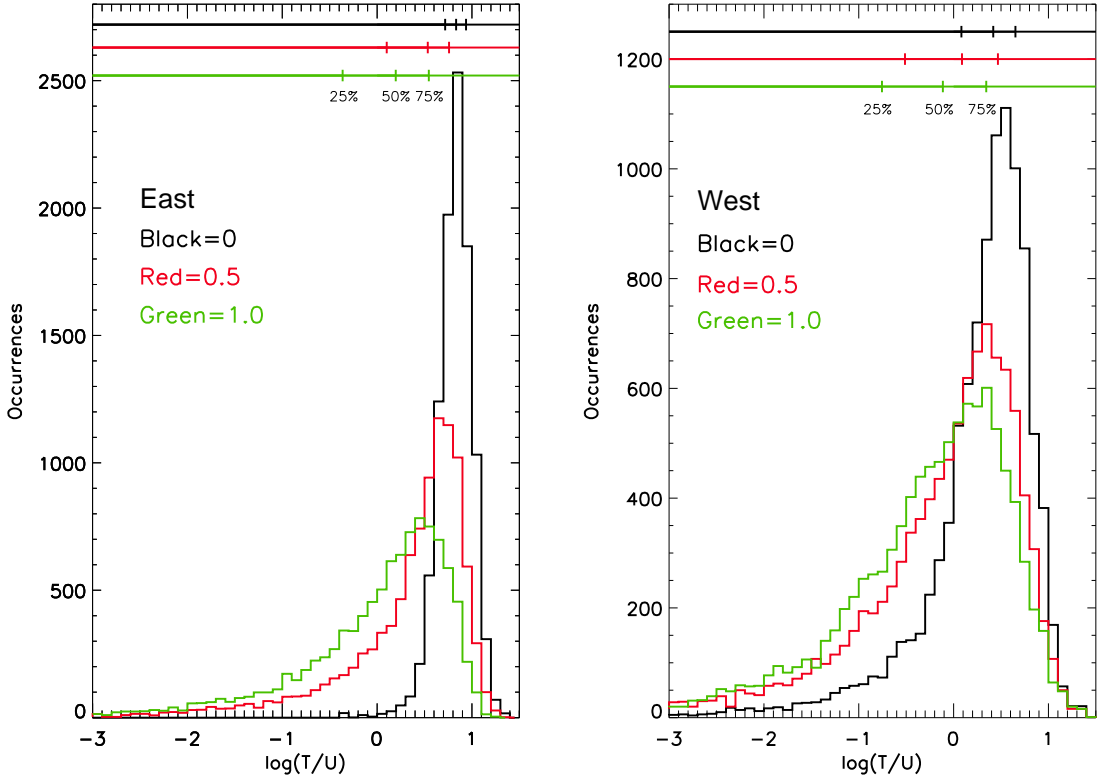}
\caption{Plot of log(T/$\mid$U$\mid$) for east (left) and west (right) sub-clusters from the posterior probability distribution accounting for a changing binary fraction. The black histogram for a binary fraction of 0, the red for a binary fraction of 0.5, and the green for a binary fraction of 1.0. The horizontal lines the top are marked for quartile positions in the distributions. In both cases, the peak log(T/$\mid$U$\mid$) evolves to a flatter and more bound distribution as the binary fraction increases.}
\label{fig:tudist}
\end{figure*}

\begin{figure*}[ht]
\centering
\plotone{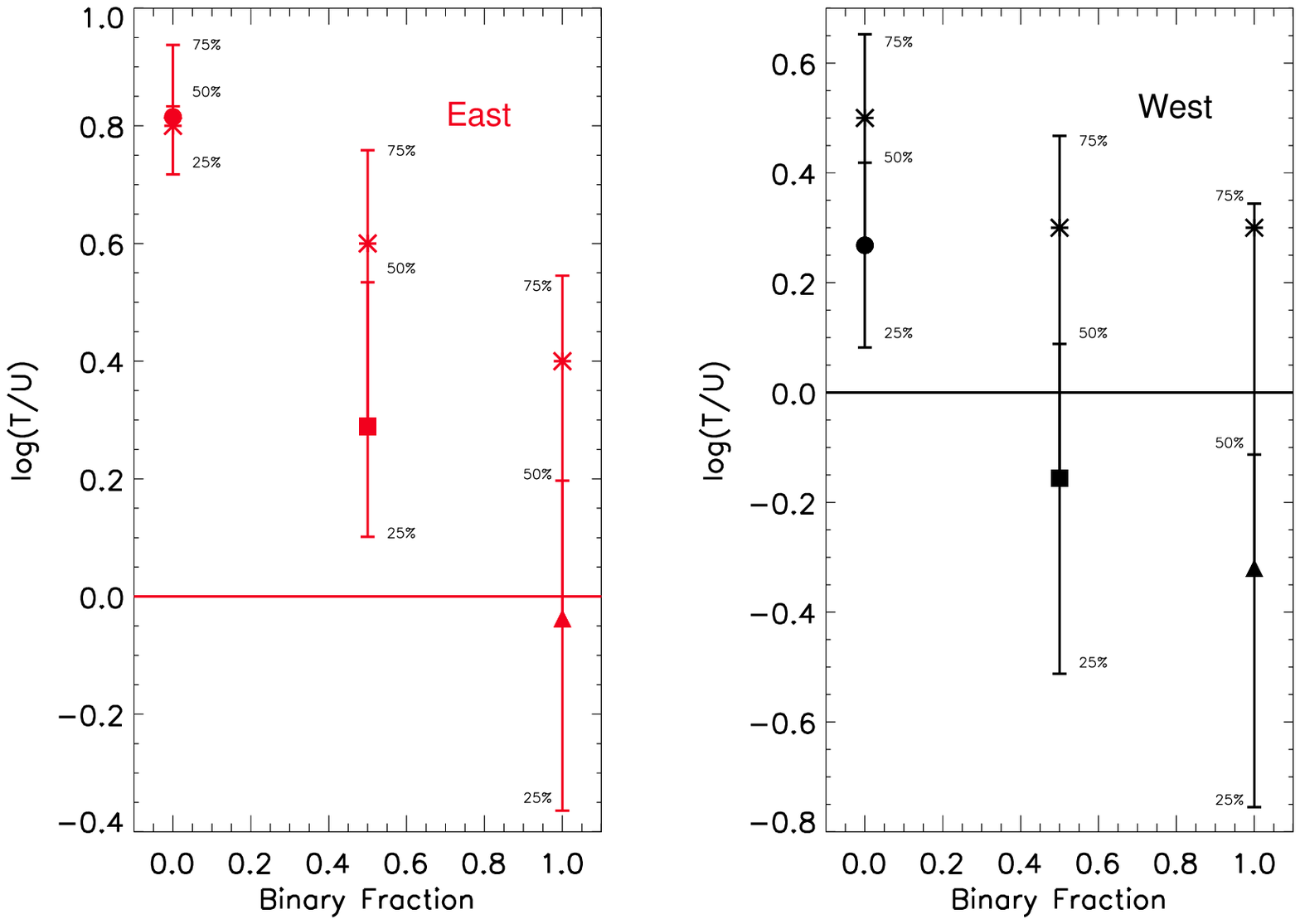}
\caption{Plot of log(T/$\mid$U$\mid$) for both sub-clusters. The eastern sub-cluster is represented in red (left) and the west sub-cluster is represented in black (right) as a function of the adopted binary fraction. The horizontal line indicates log(T/$\mid$U$\mid$) = 0.0 this is the dividing line between a bound and unbound cluster. The circles, squares, and triangles are the mean value for each distribution in Figure 10 and the overlaid asterisks are the values at the peaks of the distributions in Figure 10. The quartiles positions are indicated that correspond to those in Figure 10.}
\label{fig:bound}
\end{figure*}

\section{Discussion} 

After fitting the sizes and densities with the empirical King 1962 model and constraining the kinematics of the two sub-clusters, we discuss their possible early conditions, evolutionary paths up to their current state, and their potential future fates. 

\subsection{The Possible Early Conditions and Evolutionary Paths of the Eastern Sub-Cluster}
\label{sec:east_early}

There is clear evidence that the east sub-cluster has undergone gas expulsion due to the radiation and winds of the O7 star HD 217086. The 2MASS extinction maps show that the gas measured in the east is concentrated around the edge of the sub-cluster and mostly dispersed in the center of the cluster \citep{gutermuth2011b,allen2012}. With a current number of members and gas mass as observed today, the current star formation efficiency (SFE) is $\sim$73\%; the stellar mass dominates the gravitational potential of this sub-cluster. 

There are several reasons the east is in a state of expansion due to the gas dispersal. First, simulations of clusters after gas dispersal show that clusters expand significantly, with the ratio of the final to initial radii equal to the ratio of the initial to final stellar mass in the case the cluster remains bound \citep{baumgardt2007,moeckel2010}. An expansion factor of five with a SFE $\sim$0.2 can lead to a bound cluster if the gas dispersal timescale is slow enough, i.e., several crossing times of the cluster. Second, the circular symmetry of the cluster is unlikely due to relaxation. The crossing time at the current radius is 

\begin{equation}
	t_{cross} = R_{core} / \sigma_{1D}
\end{equation}
\noindent 

0.74$\pm$0.16 Myr. This implies $\sim$4 crossings have occurred since the formation of the cluster assuming an age of 3-4 Myr. To determine the relaxation time, we use the two adopted disk fractions and 

\begin{equation}
	t_{relax} = N t_{cross} / 6 ln(N/2),
\end{equation}
\noindent 

where N is the number of stars within the eastern sub-cluster radius. The relaxation time of the east is 16.6$\pm$3.6 Myr and 14.1$\pm$3.1 Myr using 809 and 664 as N, respectively. This indicates that at its current size, the east does not have time to dynamically relax and its circular symmetry may instead come from the expansion of the cluster from a more compact, potentially more irregular, structure. 

A comparison of the size and density of the eastern sub-cluster to younger clusters shows further evidence for expansion. The east has a core radius of 1.36$\pm$0.30 pc and a peak stellar surface density of 521 and 428 stars pc$^{-2}$. In comparison, clusters at an early stage in their gas dispersal phase have much smaller sizes and higher densities. We used a combination of $Spitzer$ and $Chandra$ surveys \citep{megeath2016}, to perform the same King model fits to the ONC and the NGC 2024 cluster in the Orion B cloud, both of which still have significant amounts of molecular gas and contain massive stars. The ONC has a core radius of 0.04462$\pm$0.00003 pc and a central peak stellar density of 1.12$\times$10$^{6}$ stars pc$^{-2}$. NGC 2024 has a core radius of 0.0305$\pm$0.0038 pc and a central peak stellar density of 5.47$\times$10$^{4}$ stars pc$^{-2}$. These two clusters bracket the number of member stars contained in the east of Cep OB3b, NGC 2024 has $\sim 400$ dusty YSOs and the ONC has 3000 dusty YSOs, although the lower stellar density regions of this cluster extend along a 10 pc filament \citep{megeath2016}. 

More recently, \citet{kuhn2018} found direct evidence in the proper motions from Gaia DR2 that the eastern sub-cluster, which is referred to as Cepheus B in their paper, is undergoing expansion. They measure parallaxes from 481 members and find a positive radial gradient and significant ($>$ 3$\sigma$) expansion of Cep B. 

Estimating the initial size of the cluster before expansion requires several assumptions. If we assume it began in virial equilibrium (2T=U), it had r$^{-2}$ density distribution and that the gas and stars have the same spatial distribution, then the outer radius of the cluster is given by R = GM / 3$\sigma^{2}_{1D}$ \citep{maclaren1988}, where M is the total mass. The stellar mass, determined by using 0.5 M$_{\odot}$ as the average mass, for the 809 and 664 members of the east are 405~M$_{\odot}$ and 332~M$_{\odot}$, respectively. We adopt a total SFE of 0.2 at the onset of gas dispersal. At that point, 20\% of the mass is in stars, and 80\% is in gas. Using SFE = M$_{stars}$ / M$_{stars}$ + M$_{gas}$ and solving for M$_{gas}$ results in 1620 M$_{\odot}$ and 1328 M$_{\odot}$, respectively. The total masses are 2025 M$_{\odot}$ and 1660 M$_{\odot}$, respectively. Assuming a constant velocity dispersion of 2.2~km s$^{-1}$, which is the expectation value for the eastern sub-cluster for a binary fraction of 0.5, we obtain an initial outer radius of 0.58~pc and 0.49~pc, respectively. The sub-cluster radius is 2.32~pc. If we use 2.2~km s$^{-1}$ as the velocity, then the eastern sub-cluster has been expanding for 0.76 Myrs or 0.80 Myrs, respectively. Note that the assumptions made are very simplistic. In particular, we would expect the velocity dispersion to decrease with time \citep{moeckel2010}. In this case, the eastern sub-cluster would have formed in a more compact, dense configuration with a smaller initial radius, which is consistent with observations of younger clusters and from simulations. Furthermore, the initial SFE may range from 0.1 to 0.3 as observed in young, embedded clusters, \citep{megeath2016}, and the initial conditions may not be virial \citep[e.g.,][]{farias2018}. Note that the value of the SFE can depend on the spatial scale on which it is measured. The inner regions of a protocluster will have a higher SFE value than the edges \citep[see Fig 13 in][]{parmentier2013}.

Although most of the gas has been dispersed, star formation is continuing at the edge of the cluster. \citet{getman2009} proposed that the O7 star is creating a radiative driven implosion (RDI) in the rim of the associated molecular clump resulting in continued star formation along the edge of the east see Figure 4 in \citet{allen2012}. Although such an implosion can produce a velocity shift, our results do not detect RV gradient with respect to the location the O star, however, the gradient could be perpendicular to our line of sight and undetectable from RV motions. Given the small number of protostars in this RDI \citep{allen2012}, this implosion may only contribute a small fraction of the cluster stars.  

It seems clear that the east has expanded since formation; however, it is unclear how many of the stars form a bound cluster. Simulations and theoretical analyses show that clusters where the kinetic energy exceeds the potential energy can still form bound clusters, although at a smaller efficiency. For an assumed star formation efficiency of 0.25 and a gas dispersal time of 1 Myr, or two crossing times, \citet{baumgardt2007} estimate roughly 55\% of the stars may remain bound. Note that \citet{baumgardt2007} assume the radial density profiles of the stars and of the residual gas in an embedded cluster have the same shape. If the density profile of the stars is steeper than the profile of the gas, which is likely for the inner regions, then the likelihood of the formation of a bound cluster increases substantially \citep[see][]{adams2000,parmentier2013,shukirgaliyev2017}.

A key parameter is T/$\mid$U$\mid$ known as the virial ratio (Q). \citet{farias2018},  ``based on the value of Q just after gas dispersal", estimates the fraction of stars that remains bound after gas dispersal. They compare simple models and simulations to show that the bound number strongly depends on the initial post gas dispersal virial ratio. We expect the value of T/$\mid$U$\mid$ to slowly increase as the cluster expands. For a binary fraction of 0.5, the expectation value of the velocity dispersion for log(T/$\mid$U$\mid$)$\sim$0.3; for this value of Q, $\sim$35\% of the stars will remain bound. At the 25th percentile the log(T/$\mid$U$\mid$) value is 0.1, which corresponds to 55\% of the members remaining bound and at the 75th percentile, the log(T/$\mid$U$\mid$) value is 0.76, which corresponds to 10\% of the members will remain bound. If the velocity dispersion is closer to those inferred for a binary fraction of 0 then 10\% or less will remain bound. If the binary fraction is 1, then up to 95\% of the members remain bound in the 25th percentile and as few as 20\% members will remain bound in the 75th percentile. \citet{kuhn2018} found a velocity dispersion of 1.9$\pm$0.2 km s$^{-1}$ and that the cluster is expanding radially. This is similar to the expectation value with a velocity dispersion of 2.2~km s$^{-1}$ for a binary fraction of 0.5, which is the binary fraction that is most consistent with that of solar type stars \citep{duchene2013}. Although a broad range of outcomes are allowed, we favor where approximately a third of the stars remain bound.

\begin{figure*}[ht]
\centering
\plotone{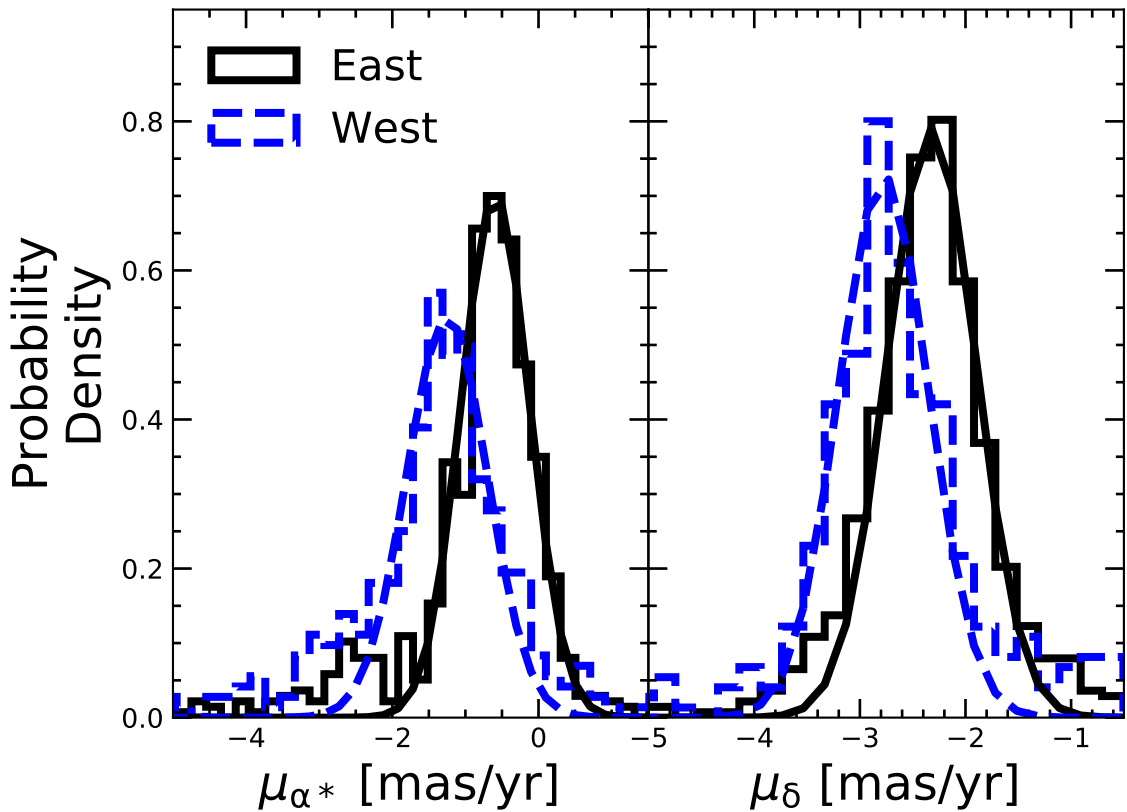}
\caption{Histograms of the member proper motions in RA (left) and Dec (right) for the east (solid black) and west (dashed blue) sub-clusters. The bin sizes are 0.4 mas yr$^{-1}$. The peaks of the probability densities show that the two sub-clusters have different values of proper motion components: $\mu$$_{\alpha}$=-0.59$\pm$0.02 (mas yr$^{-1}$) and $\mu$$_{\delta}$=-2.32$\pm$0.02 (mas yr$^{-1}$) for the east and $\mu$$_{\alpha}$=-1.25$\pm$0.03 (mas yr$^{-1}$) and $\mu$$_{\delta}$=-2.78$\pm$0.02 (mas yr$^{-1}$) for the west. The gaussian fits to the distributions are shown.} 
\label{fig:propermotionhisto}
\end{figure*}

\subsection{The Possible Early Conditions and Evolutionary Paths of the Western Sub-Cluster}

The smaller western sub-cluster remains more embedded than the eastern sub-cluster. We calculate a current SFE between 0.23 and 0.26 depending on the disk fraction, which is much higher than the eastern sub-cluster. Even though it is similar to SFEs of embedded clusters, the gas is offset to the south of the center of the sub-cluster, which appears to have at least partially dispersed gas \citep{allen2012}. Additional evidence that it is not as dynamically evolved as the eastern sub-cluster is its elongated morphology and smaller core size (0.5$\pm$0.3 pc). 

The lesser degree of dynamical evolution in the west does not necessarily imply a younger age. The west contains one B3 and three B5 stars that do not have the UV radiation of the O7 star in the eastern sub-cluster to clear natal gas as quickly. Calculating a crossing time as in \S \ref{sec:east_early}, yields t$_{crossing}$ = 1.45$\pm$0.31 Myr. The west has experienced $\sim$3 crossings since the formation of the cluster, similar to the eastern sub-cluster, assuming an age of 3-4 Myr. The relaxation time is $\sim$19 - 22 Myr for the two disk fractions. This indicates the west is not dynamically relaxed and could maintain its elongated morphology. The core radius is larger than the ONC and NGC 2024 (see \S \ref{sec:east_early}) and the surface densities (220 and 182 stars pc$^{-2}$) are lower suggesting that the core of the sub-cluster has expanded. 

As the western sub-cluster evolves it is unclear how quickly the remaining natal gas will disperse and how the virial ratio will change as a result. It is a reasonable assumption that T/$\mid$U$\mid$ will increase as it expands, but by how much remains unclear. Our range of log(T/$\mid$U$\mid$) values (Figure 11) are consistent with a cluster that is currently virialized, although with large error bars.   

If this virial ratio is maintained through gas dispersal then we can again compare our log(T/$\mid$U$\mid$) values to Figure 7 in \citet{farias2018} to determine the fraction of stars that will remain bound as in \S \ref{sec:east_early}. For a binary fraction of 0.5, the expectation value of log(T/$\mid$U$\mid$) is -0.16, which yields $\sim$75\% of the stars will remain bound. At the 25th and the 75th percentiles the log(T/$\mid$U$\mid$) value is -0.5 and 0.47, which corresponds to 98\% and 20\% members remaining bound, respectively. If the velocity dispersion is closer to those inferred for a binary fraction of 0 then the expectation value of log(T/$\mid$U$\mid$) is 0.27 resulting in 30\% of the stars remaining bound. For the 25th and 75th percentiles as many as 50\% and as few as 10\% of the members will remain bound, respectivley. If the binary fraction is 1 the expectation value results in 95\% of the stars to remain bound. If we take the 25th and 75th percentiles the number of remaining members will be 100\% and 30\%, respectively. Although a broad range of outcomes are allowed, we favor where approximately 75\% of the stars remain bound. \citet{baumgardt2007} found that a combination of very slow gas expulsion and a very weak external tidal field with an initial SFE of 33\% can produce a bound cluster with 90\% of stars remaining bound. The current SFE is 26\% and it is possible that the west initially had a SFE of $\sim$30\%. If the timescale of gas dispersal is prolonged, which is the likely case in the west, then the fraction of bound stars may be higher \citep{baumgardt2007}. 

\begin{figure*}[ht]
\centering
\plotone{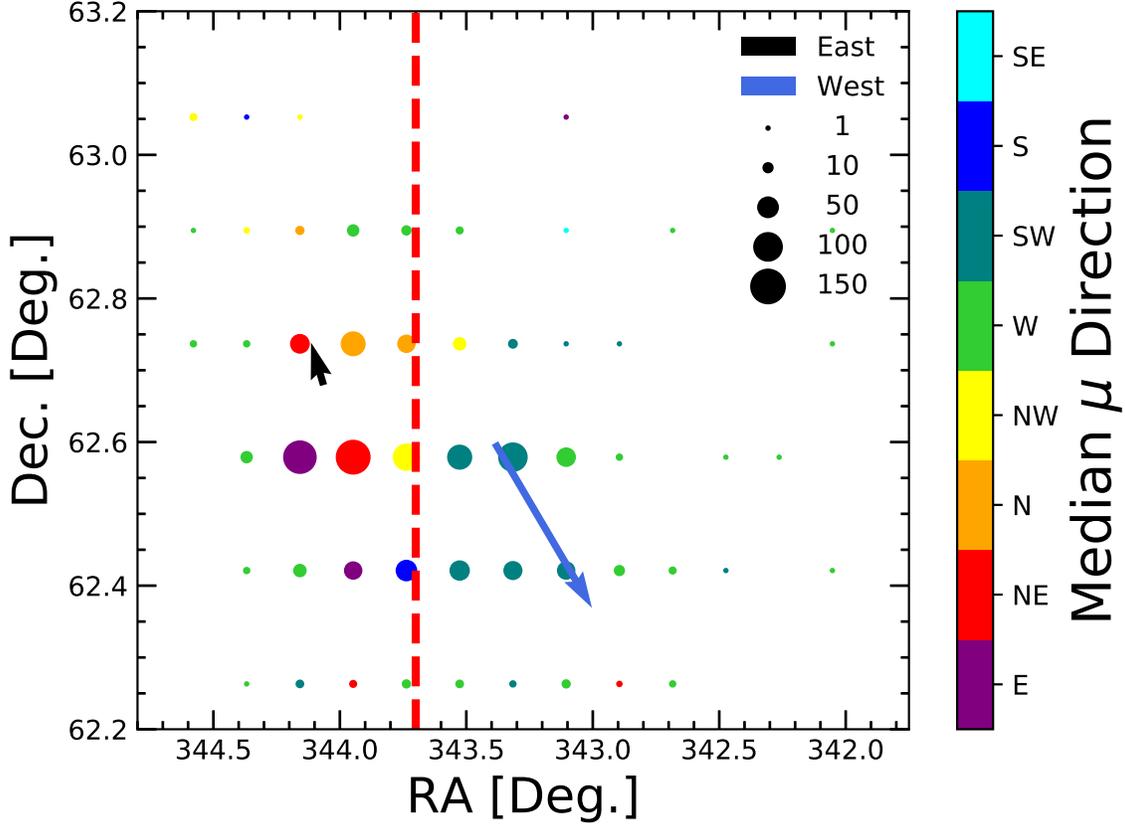}
\caption{The total proper motions of the east (black arrow) and west (blue arrow) sub-clusters with the average proper motion of the total cluster removed demonstrating the sub-clusters will not merge but remain separate. The red, vertical dashed line indicates the separation in RA of the two sub-clusters. The colored dots correspond to the median parallax direction at different points in Cep OB3b with the size increasing as the number of stars in the bin agree with the direction of motion. The eastern sub-cluster has a total proper motion of 2.34$\pm$0.02 mas yr$^{-1}$. The western sub-cluster has a total proper motion of 2.86$\pm$0.03 mas yr$^{-1}$. The differential velocity exceeds the escape velocity from the more massive east.} 
\label{fig:propermotioncolor}
\end{figure*}

\subsection{Fate of Cep OB3b}

Our values for the virial ratios are consistent with Cep OB3b forming two bound sub-clusters, each with $\sim$300 stars. These two sub-clusters would be found within an expanding association of stars from the lower density halo surrounding both of the sub-clusters, as well as, the members ejected from the sub-clusters. The associations would have more stars than the clusters combined. 

To determine whether the two sub-clusters may merge into a single cluster, we compare the escape velocity to the relative motions of the two sub-clusters. The escape velocity is v$_{esc}$=(2GM$_{tot}$/R), where M$_{tot}$ is the combined stellar mass of the east and west. We use the most optimistic masses of stars, i.e., (809 $+$ 501) $\times$ 0.5 M$_{\odot}$ and a separation of the sub-clusters of 5 pc as R. This results in an escape velocity of 1.06 km s$^{-1}$. The total proper motion of Cep OB3b is subtracted off the proper motions of the sub-clusters. The east is moving at -0.47 km s$^{-1}$ and the west is moving at 1.55 km s$^{-1}$ for a difference in 2.02 km s$^{-1}$. Considering that the masses of the sub-clusters will be lower after gas dispersal and that the actual separation is likely to be larger than the projected separation, this escape velocity should be considered as an upper limit. Thus, the two sub-clusters are moving away from each other at a velocity that exceeds the escape velocity and only increase in separation forming a double cluster.

This result seems exceptional in the light that $\sim$7\% of embedded clusters form open clusters \citep{lada2003}. This might be due to unusual conditions in the Cep OB3b cluster. Alternatively there might be other factors that contribute to the dissipation of clusters, \citep[see][]{moeckel2012}, particularly when a realistic range of masses are used. Stellar dynamical effects come into play when binaries and a realistic range of masses are used \citep[][]{moeckel2012}. Alternatively, the tidal forces between the clusters, and between the clusters and the molecular clouds may disrupt the clusters or strip members. It should also be noted that the large uncertainties in the velocity dispersions translate into large uncertainties in the dynamical outcomes. If both sub-clusters form bound clusters then the study of Cep OB3b will help establish the cluster properties and environmental factors which produce bound clusters. 

\begin{figure*}[ht]
\centering
\plotone{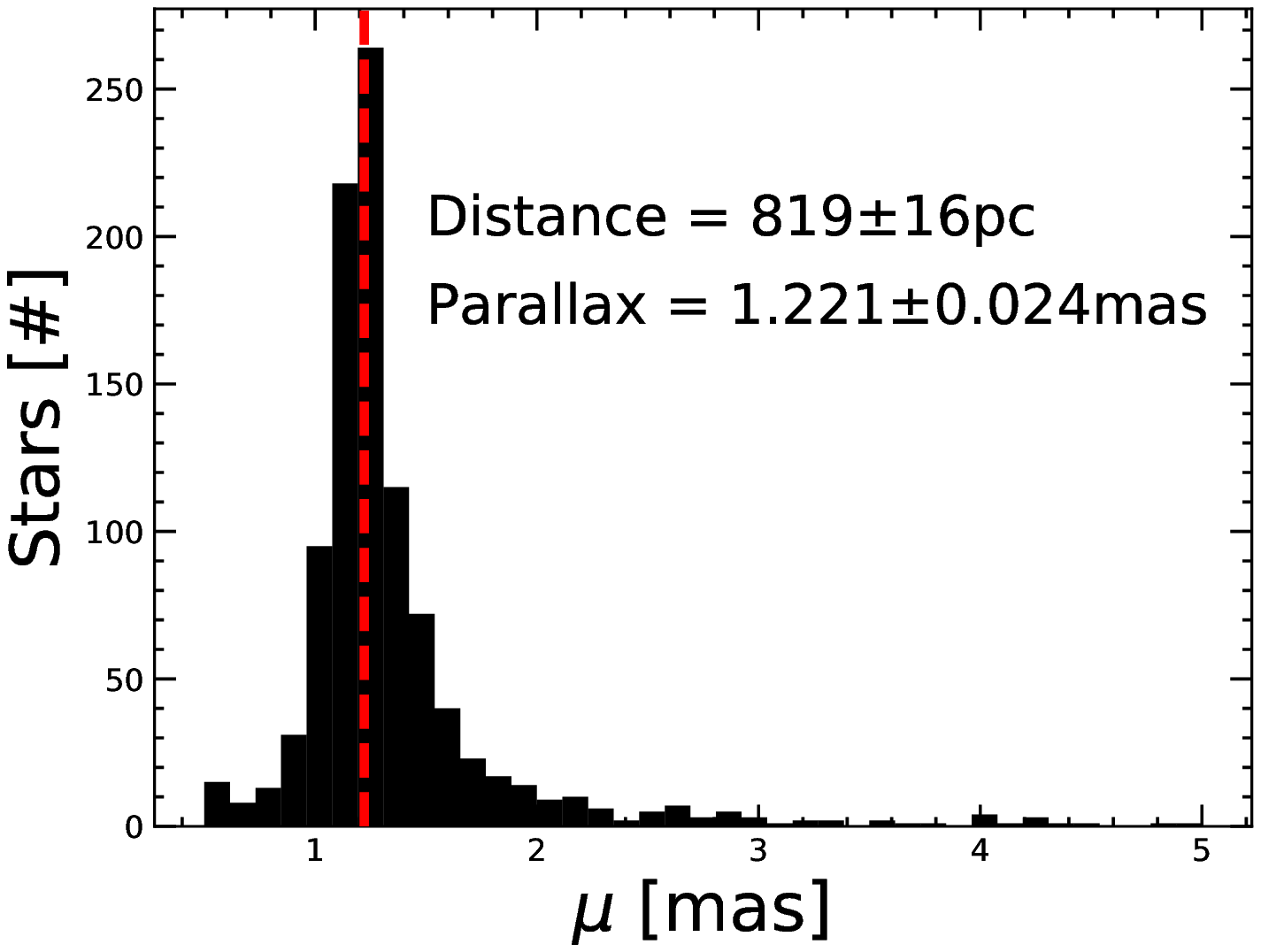}
\caption{A histogram of objects toward Cep OB3b with parallaxes measured by Gaia in DR2. The median parallax is marked by the dashed, vertical red line yielding a distance of 819$\pm$16 pc. We adopt this distance for Cep OB3b.} 
\label{fig:cepdistance}
\end{figure*}

\subsection{The Effect of Binarity on the Velocity Dispersions of Young Clusters}

One of the results of our analysis is that the adopted binary fraction strongly influences the PDFs for the cluster velocity dispersions, as demonstrated in Figure 8. This importance can be further demonstrated by a simple comparison of the expected velocity dispersion due to binarity compared to that from clusters. We compare these by approximating the ratio of the velocity dispersion from cluster motions to those from binary motions.

The 1-D velocity dispersion of a barely bound ($U = T$) cluster is given by

\begin{equation}
  \sigma^2_{1D} = \alpha \frac{G M}{R},
\end{equation}

\noindent
where $M$ is the total mass of the cluster, $R$ is the core radius of the cluster, and $\alpha$ depends on the radial density distribution of the spherically symmetric cluster. We adopt the value of $\alpha = 2/5$, which is that for a uniform density cluster. For a cluster, the observed velocity dispersion due to the motions of binaries is approximately 

\begin{equation}
   \sigma^2_{bin} = 0.5~f_{bin}~\beta \frac{G m}{r},
\end{equation}

\noindent
where $m$ is the combined mass of the system, $r$ is the typical semi-major axis, $f_{bin}$ is the binary fraction, and $\beta$ takes into account the effect of orbital inclinations and orbital phases on the line of sight velocity. The 0.5 term assumes that the reduced mass is half the combined mass. We adopt $\beta = 2/3$, $m = 1.5 $~M$_{\odot}$, and a semi-major axis of 45~AU; this is the peak of the lognormal semi-major axis distribution for solar mass, field stars \citep{duchene2013}. Throughout the following analysis, we assume $f_{bin} = 0.5$. The adopted system mass of 1.5~M$_{\odot}$ assumes that the stars with measured velocity dispersions have masses of 0.75~$M_{\odot}$ each.  

To compare the magnitudes of the binary motions and motions within a cluster, we consider the ratio of the two velocities dispersions, which is given by

\begin{equation}
\frac{\sigma^2_{bin}}{\sigma^2_{1D}} = 0.5 f_{bin} \left(\frac{\beta}{\alpha}\right) \left(\frac{m_{bin}}{N_{\star} m_{\star}}\right)\left(\frac{R}{r}\right),
\label{eqn:ratio}
\end{equation}

\noindent
where $N_{\star}$ is the total number of stellar systems (including single and multiple stars) in the cluster and $m_{\star}$ is the average mass of the cluster members.  The adopted mass of the cluster members is $m_{bin} = 0.75 (1+f_{bin})$~M$_{\odot}$, where the average mass an individual star with a measured velocity is assumed to be 0.75~M$_{\odot}$. For a cluster with $N_{\star} = 500$, we find that for the ratio of velocity dispersions to be equal to or less than one (i.e. that the cluster motions are equal to or exceed those of the binaries), the cluster radius must be $\le 0.2$~pc.  This ratio increases linearly with the number of stars. Thus, for all but the youngest, most compact clusters, the motions of the binaries dominate the observed velocity dispersion. 

For embedded clusters, the dominant mass is that of the molecular gas. This requires us to include the SFE in Eqn~\ref{eqn:ratio}; the SFE is the ratio of the mass in stars over the combined stellar and gas mass. Specifically, the value of $N_{\star} m_{\star}$ must be replaced with $N_{\star} m_{\star}/SFE$. If we adopt $SFE = 0.2$, a common value for young clusters at earlier phases of gas dispersal \citep[e.g., ][]{megeath2016}, then the velocity dispersion ratio is unity or less for $R \le 1$~pc. Thus, in the case of young embedded clusters, which have typical radii of 0.5~pc, the contribution cluster motions can exceed or equal those due to binaries due to the effect of the gas mass on the cluster motions. As the gas is dispersed and the cluster expands, binaries will begin to dominate the observed distribution if it remains in virial equilibrium while expanding.   

\section{Summary}

We present a study of the spatial structure and kinematics of Cep OB3b, a young cluster currently in a late stage of gas dispersal. Cep OB3b is one of the closest examples of a $\sim$3 - 5 Myr, large ($\sim$3000 total members) cluster. The cluster is broken into two sub-clusters: the east that contains an O7 star and west, which has several B-stars. Using the combined sample of members identified with IR-excesses or X-ray detections \citep{allen2012}, the structure, density structures, and deviation from circular symmetry of the two sub-clusters are determined from fitting the empirical density law of King (1962). We present new RV measurements from Hectoschelle of 499 stars; 109 of these are likely confirmed members of Cep OB3b with youth indicators. After excluding potential binaries and applying a R cut, 3$\sigma$-clipping, and sub-cluster radius cut to the data, we are left with 52 stars in our RV analysis.

\begin{itemize}
\item{We derive the distance to CepOB3b using known members found in the Gaia DR2 catalog. The distance is 819$\pm$16 pc.}

\item{Fits to the empirical \citet{king1962} models yield the properties of the cluster. For the eastern sub-cluster, we find a core radius of 1.36$\pm$0.30 pc and a peak density of 428 to 521 stars pc$^{-2}$, depending on the adopted disk fraction. For the western sub-cluster, we find a core radius of 0.52$\pm$0.11 pc and a peak density of 284 and 342 stars pc$^{-2}$. The sub-cluster radii of the sub-clusters are 2.32$\pm$0.19 pc in the east, with 664 to 809 stars within this border and 3.1$\pm$1.0 pc in the west, with 332 to 402 stars within this border.}

\item{We have determined the posterior probability distribution for the velocity dispersion of each sub-cluster. This analysis includes a contribution due to binaries for an assumed binary fraction. The peaks of the distributions in the east are 2.8 km s$^{-1}$, 2.2 km s$^{-1}$, and 1.15 km s$^{-1}$ for a binary fractions of 0, 0.5, and 1, respectively. The expectation values are $2.83^{+0.24}_{-0.25}$ km s$^{-1}$, $1.91^{+0.50}_{-0.42}$ km s$^{-1}$, and $1.40^{+0.40}_{-0.52}$ km s$^{-1}$ for binary fractions of 0, 0.5, and 1, respectively. The west has velocity dispersion peaks at 1.5 km s$^{-1}$, 0.5 km s$^{-1}$, and 0.3 km s$^{-1}$  with expectation values of $1.49^{+0.30}_{-0.34}$ km s$^{-1}$, $1.10^{+0.30}_{-0.44}$ km s$^{-1}$, and $0.95^{+0.24}_{-0.46}$ km s$^{-1}$ for a binary fractions of 0, 0.5, and 1, respectively. Using Gaia DR2 proper motions, \citet{kuhn2018} find a velocity dispersion of 1.9$\pm$0.2 km s$^{-1}$ for the east.}

\item{A comparison of the eastern sub-cluster with the NGC 2024 and ONC cluster indicates that this cluster has a much larger core radius and lower central density.  It also is circularly symmetric. This is evidence that this cluster has undergone significant expansion.  Although the western sub-cluster is more compact than the eastern and shows a significant asymmetry, it still has a larger core radius and lower central density than the two Orion clusters.  This sub-cluster also appears to undergone some degree of expansion.}

\item{The inferred ratio of the kinetic to potential energy of the eastern sub-cluster, T/$\mid$U$\mid$, shows that this ratio depends strongly on the adopted binary fraction for the stars. For a binary fraction of 0 and 0.5, the log(T/$\mid$U$\mid$) $>$ 0, suggesting the sub-cluster will undergo expansion. For a binary fraction of 1, which is unlikely, there is a moderate probably that the cluster could be bound.  We conclude that this sub-cluster most likely have a ratio greater than 1 and is expanding.}

\item{A similar analysis for the western sub-cluster show log(T/$\mid$U$\mid$)$\sim$0 for binary fractions of 0.5 and 1, but log(T/$\mid$U$\mid$) $> 0$ for a binary fraction of 0. We conclude that this sub-cluster is close to bound and may be virialized.}

\item{Accounting for unresolved binaries is important to accurately probe the dynamical properties of young clusters, particularly after gas dispersal. The binary motions can dominate the motions of the cluster and this is particularly important when limited to solar mass stars with a high fraction of multiplicity.}

\item{For the binary fraction of 0.5, we find that the eastern sub-cluster is more likely in a state of expansion. Comparisons to nbody simulations suggest 35\% of the member stars may remain to form a bound cluster. In contrast, the western sub-cluster may be near virial equilibrium where close to 75\% of the members remain bound. These two different outcomes may be driven by the rapid of gas dispersal of gas in the eastern sub-cluster due to the presence of an O7 star. In contrast, the western-sub cluster, which still contains a significant mass of gas, only contains massive stars of spectral type B3 or later.}

\item{A likely outcome is the two sub-clusters will form bound clusters with ~300 stars. An analysis of the bulk proper motion of the two sub-clusters using GAIA DR2 shows that the sub-clusters are moving away from each other at $\sim$2 km s$^{-1}$, and they are not bound. Thus, Cep OB3b may be forming two, independent bound clusters. Given that only $\sim 7$\% of embedded clusters survive to form bound clusters, this is a very rare outcome, suggesting that the physical conditions in Cep OB3b are high conducive to bound cluster formation. Alternatively, other factors which have been ignored in our analysis, such as internal dynamics of clusters or tidal forces may play an important role in the disruption of clusters; if this is the case, the nascent bound clusters in Cep OB3b may still be disrupted.}

\end{itemize}

\section{Appendix}

\subsection{HK Emission}

Ca II H and K line emission is an indicator of youth and a way of identifying diskless pre-ms stars that do not have detectable X-ray emission. Emission in the H\&K lines (3968.5~\AA~and 3933.7~\AA,~respectively) in a spectrum may indicate magnetic activity in the chromosphere. As stars contract onto the main sequence their magnetic fields can be stronger than when they reach the main sequence. The increased magnetic field may be due to rapid rotation of stars in the pre-ms phase \citep{johnskrull1999}. We used the Hectospec data to determine if any of our objects had H\&K emission. When we observed emission in the line core, the object was marked to have H\&K emission. Fifteen objects with Ca II H\&K line emission in our sample had no previous indications of youth as well as 25 objects that had at least one youth indicator already.

As seen in Figure 15, most (83\%) objects with HK emission have RVs within 3$\sigma$ of the average RV of the cluster. Whereas 46\% of the objects without youth indicators are within 3$\sigma$ of the average RV of the cluster. The outliers with HK emission may be binaries that we simply haven't detected or may indicate that $\sim$20\% of the objects identified by H\&K emission may be contaminants.

\begin{figure*}
\centering
\plotone{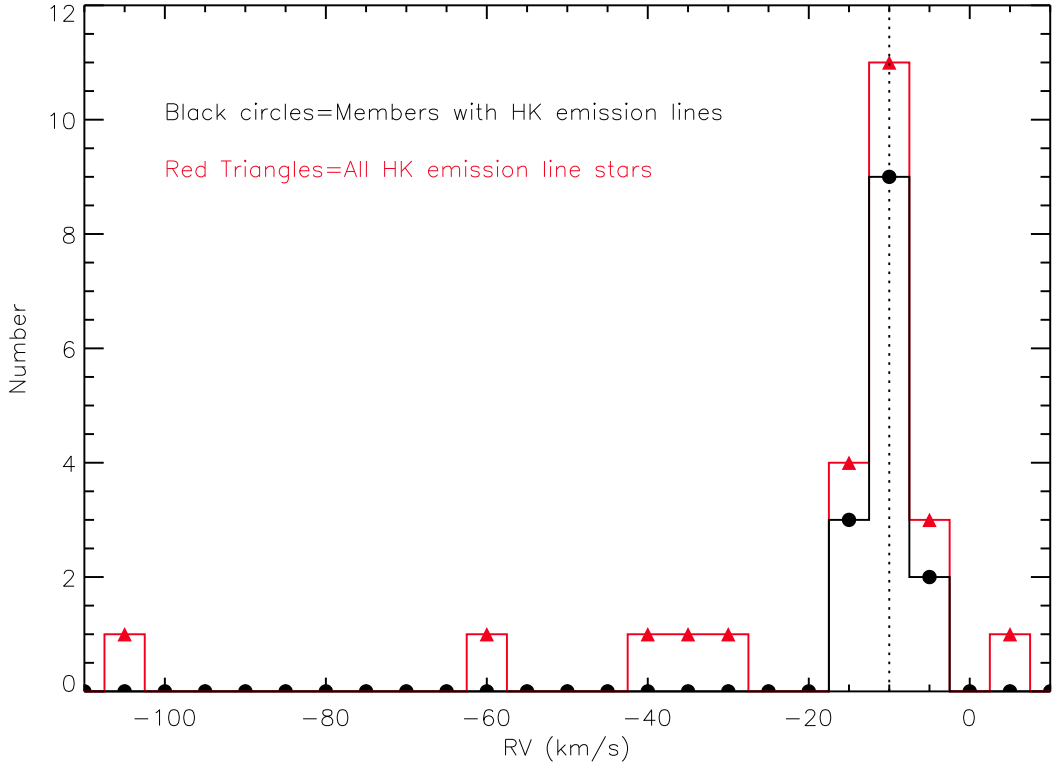}
\caption{RVs of stars histogram marked by with HK emission lines and R values larger than 5.0. The histograms are marked by the black circles is for objects with youth indicators and the red triangles is for all objects with HK emission lines. There is a clear peak of the objects with youth indicators at -10 km s$^{-1}$ (dotted line) that corresponds to the V$_{lsr}$ peak of Cep OB3b.} 
\label{fig:HKhisto}
\end{figure*}

\vspace{12 pt}
\indent \textbf{Acknowledgements}

We wish to thank the anonymous referee for excellent suggestions that improved the quality of this manuscript. The authors are also grateful to M. Kounkel for useful discussions and comments throughout the progress of
this project. This work was supported by the NSF grant AST-1009564 to S.T. Megeath. RAG gratefully acknowledges funding support from NASA ADAP grants NNX11AD14G, NNX15AF05G, and NNX17AF24G, NASA JPL/Caltech contract 1489384, and NSF grant AST 1636621 in support of TolTEC, the next generation mm-wave camera for LMT. 
Data presented herein were obtained at the MMT Observatory at Fred Lawrence Whipple Observatory on Mount Hopkins, AZ USA. 
This work made use of the SIMBAD database, the Vizier database, the NASA
Astrophysics Data System, and the data products from the Two Micron All Sky Survey, which is a joint
project of the University of Massachusetts and the Infrared Processing and Analysis Center/California
Institute of Technology, funded by the National Aeronautics and Space Administration and the NSF.

This work has made use of data from the European Space Agency (ESA) mission
{\it Gaia} (\url{https://www.cosmos.esa.int/gaia}), processed by the {\it Gaia}
Data Processing and Analysis Consortium (DPAC, \url{https://www.cosmos.esa.int/web/gaia/dpac/consortium}). Funding for the DPAC
has been provided by national institutions, in particular the institutions
participating in the {\it Gaia} Multilateral Agreement.

\bibliography{bibcepob3b}

\newpage 

\begin{deluxetable*}{cccccccc}
\tablewidth{0pt}
\tablecaption{Hectoschelle Observations
 \label{tbl:1}}
\tablehead{
\colhead{UT Date      }  & \colhead{R.A.    } & \colhead{Dec.     } & \colhead{Exposures} & \colhead{Seeing} & \colhead{Airmass} & \colhead{$\#$ of targets} \\
\colhead{$ $} & \colhead{(J2000)} & \colhead{(J2000)} & \colhead{($\#$ $\times$ seconds)} & \colhead{(arcsec)} & \colhead{$ $} & \colhead{$ $}}
\startdata
2009 Oct 26 & 22 55 12.36 & +62 40 11.64  & 4 $\times$ 1800 & 1.97 & 1.26 & 120 \\
2009 Nov 03 & 22 55 09.12 & +62 39 07.56 & 4 $\times$ 1800 & 1.76 & 1.23 & 123 \\
2010 Nov 25 & 22 55 07.86 & +62 42 16.02 & 4 $\times$ 1800 & 1.83 & 1.23 & 88 \\
2011 Oct 20 & 22 55 08.04 & +62 41 35.52 & 4 $\times$ 1800 & 0.72 & 1.25 & 153 \\
2013 June 25 & 22 55 43.68 & +62 39 03.24 & 4 $\times$ 2100 & 1.74 & 1.53 & 77 \\
\enddata

\end{deluxetable*}

\newpage

\begin{deluxetable*}{ccccc}
\tablewidth{0pt}
\tablecaption{2009.1026 RV Results
 \label{tbl:1026}}
\tablehead{
\colhead{2MASS ID} & \colhead{RA (J2000)} & \colhead{Dec (J2000)} & \colhead{RV$\pm\sigma$ (km s$^{-1}$)} & \colhead{R Value}}
\startdata
J22554779+6251042 & 22:55:47.76 & 62:51:04.39 & -26.348$\pm$5.8031 & 0.80 \\
J22551448+6245480 & 22:55:14.47 & 62:45:47.96 & -35.166$\pm$10.3918 & 0.95 \\  
J22553253+6253096 & 22:55:32.52 & 62:53:09.48 & -22.138$\pm$4.2083 & 2.05 \\ 
J22554968+6255229 & 22:55:49.70 & 62:55:22.88 & -5.362$\pm$0.8118 & 13.36 \\
J22561275+6300232 & 22:56:12.75 & 63:00:23.22 & -69.876$\pm$5.0686 & 2.23 \\
J22550645+6241517 & 22:55:06.47 & 62:41:51.78 & 15.388$\pm$2.5402 & 4.43 \\
J22564218+6255417 & 22:56:42.21 & 62:55:41.48 & -4.436$\pm$1.0387 & 10.47 \\ 
J22564927+6255456 & 22:56:49.26 & 62:55:45.83 & -0.277$\pm$0.8332 & 13.06 \\
\enddata
\end{deluxetable*}

\newpage

\begin{deluxetable*}{ccccc}
\tablewidth{0pt}
\tablecaption{2009.1103 RV Results
\label{tbl:1103}}
\tablehead{
\colhead{2MASS ID} & \colhead{RA (J2000)} & \colhead{Dec (J2000)} & \colhead{RV$\pm\sigma$ (km s$^{-1}$)} & \colhead{R Value}}
\startdata
J22535500+6243412 & 22:53:55.00 & 62:43:41.24 & -66.967$\pm$4.0655 & 14.79 \\
J22542631+6242377 & 22:54:26.33 & 62:42:37.77 & -10.373$\pm$1.7326 & 8.47 \\
J22542993+6240335 & 22:54:29.92 & 62:40:33.57 & -8.106$\pm$7.0226 & 2.21 \\
J22535448+6244536 & 22:53:54.47 & 62:44:53.72 & -18.343$\pm$4.2483 & 1.43 \\
J22541764+6248194 & 22:54:17.64 & 62:48:19.40 & -28.771$\pm$1.5439 & 6.33 \\
J22542450+6245330 & 22:54:24.51 & 62:45:32.96 & -23.567$\pm$1.5498 & 7.31 \\
J22540417+6247032 & 22:54:04.16 & 62:47:03.23 & -16.083$\pm$3.7199 & 2.90 \\
J22543463+6242480 & 22:54:34.62 & 62:42:47.99 & -51.037$\pm$7.6444 & 12.54 \\
\enddata
\end{deluxetable*}

\newpage

\begin{deluxetable*}{ccccc}
\tablewidth{0pt}
\tablecaption{2010 RV Results
\label{tbl:2010}}
\tablehead{
\colhead{2MASS ID} & \colhead{RA (J2000)} & \colhead{Dec (J2000)} & \colhead{RV$\pm\sigma$ (km s$^{-1}$)} & \colhead{R Value}}
\startdata
J22542652+6247077 & 22:54:26.50 & 62:47:07.79 & 39.492$\pm$5.246 & 0.99 \\
J22543806+6251022 & 22:54:38.06 & 62:51:02.27 & -38.549$\pm$1.135 & 9.35 \\ 
J22542386+6255382 & 22:54:22.92 & 62:55:24.16 & 40.462$\pm$3.222 & 1.65 \\
J22543139+6251547 & 22:54:31.39 & 62:51:54.54 & 4.479$\pm$2.091 & 4.50 \\
J22542604+6255213 & 22:54:27.90 & 62:55:26.17 & 50.149$\pm$4.531 & 1.77 \\
J22545426+6254405 & 22:54:54.24 & 62:54:40.48 & -105.702$\pm$0.734 & 14.22 \\
J22551602+6255125 & 22:55:15.89 & 62:55:21.30 & -36.919$\pm$4.063 & 1.85 \\ 
J22550634+6253203 & 22:55:06.26 & 62:53:20.52 & -39.923$\pm$1.463 & 6.96 \\
\enddata
\end{deluxetable*}

\newpage

\begin{deluxetable*}{ccccc}
\tablewidth{0pt}
\tablecaption{2011 RV Results
\label{tbl:2011}}
\tablehead{
\colhead{2MASS ID} & \colhead{RA (J2000)} & \colhead{Dec (J2000)} & \colhead{RV$\pm\sigma$ (km s$^{-1}$)} & \colhead{R Value}}
\startdata
J22544381+6243128 & 22:54:43.83 & 62:43:12.75 & -31.446$\pm$2.012 & 4.81 \\
J22535448+6244536 & 22:53:54.47 & 62:44:53.72 & 44.655$\pm$4.185 & 1.05 \\
J22542553+6244554 & 22:54:25.56 & 62:44:55.54 & -39.752$\pm$1.168 & 8.70 \\
J22543560+6246312 & 22:54:35.59 & 62:46:31.16 & 9.574$\pm$4.798 & 2.31 \\
J22545800+6243449 & 22:54:58.03 & 62:43:44.95 & -30.834$\pm$1.437 & 6.59 \\
J22544253+6245173 & 22:54:42.52 & 62:45:17.37 & -75.627$\pm$0.856 & 11.72 \\
J22541822+6245397 & 22:54:18.24 & 62:45:39.83 & -121.277$\pm$0.702 & 14.38 \\ 
J22543123+6245142 & 22:54:31.25 & 62:45:14.10 & -57.970$\pm$0.688 & 14.66 \\
\enddata
\end{deluxetable*}

\newpage

\begin{deluxetable*}{ccccc}
\tablewidth{0pt}
\tablecaption{2013 RV Results
\label{tbl:2013}}
\tablehead{
\colhead{2MASS ID} & \colhead{RA (J2000)} & \colhead{Dec (J2000)} & \colhead{RV$\pm\sigma$ (km s$^{-1}$)} & \colhead{R Value}}
\startdata
J22562637+6241289 & 22:56:26.37 & 62:41:28.97 & -10.597$\pm$1.636 & 7.30 \\
J22560480+6239460 & 22:56:04.80 & 62:39:46.09 & -9.737$\pm$1.003 & 9.98 \\
J22571825+6243042 & 22:57:18.25 & 62:43:04.27 & -11.131$\pm$2.177 & 4.56 \\
J22562231+6240488 & 22:56:22:31 & 62:40:48.81 & -9.851$\pm$1.355 & 7.90 \\
J22570134+6242133 & 22:57:01.35 & 62:42:13.39 & 11.918$\pm$4.255 & 2.77 \\
J22570594+6238180 & 22:57:05.95 & 62:38:18.07 & -13.393$\pm$2.498 & 3.43 \\
J22570917+6238560 & 22:57:09.18 & 62:38:56.06 & -16.386$\pm$4.785 & 2.88 \\
J22565186+6239521 & 22:56:51.86 & 62:39:52.12 & -14.386$\pm$1.817 & 6.48 \\
\enddata
\end{deluxetable*}

\newpage

\begin{deluxetable*}{cccc}
\tablewidth{0pt}
\tablecaption{Potential Spectroscopic Binary Candidates
	\label{tbl:binaries}}
\tablehead{
\colhead{2MASS ID} & \colhead{R.A. (J2000)} & \colhead{Dec. (J2000)}  & \colhead{RV$\pm\sigma$ (km s$^{-1}$)}}

\startdata
J22553151+6253093 & 22:55:32.52 & 62:53:09.48 & -22.138$\pm$4.204\\
J22550645+6241517 & 22:55:06.47 & 62:41:51.78 & 15.388$\pm$2.253 \\
J22560723+6245358 & 22:56:07.23 & 62:45:35.76 & -15.697$\pm$6.153 \\
J22552722+6243290 & 22:55:27.23 & 62:43:28.90 & 15.196$\pm$5.353 \\
J22561571+6242509 & 22:56:15.72 & 62:42:51.08 & -12.909$\pm$2.577 \\
J22560574+6242599 & 22:56:05.75 & 62:42:59.74 & 6.907$\pm$2.839 \\
J22570068+6245485 & 22:57:00.70 & 62:45:48.44 & -20.451$\pm$10.687 \\
J22563906+6245092 & 22:56:39.04 & 62:45:09.22 & 36.562$\pm$1.558 \\
\enddata
\end{deluxetable*} 

\newpage 

\begin{deluxetable*}{ccccccccc}
\tablewidth{0pt}
\tablecaption{King 1962 Fit Results
 \label{tbl:kingfitresults}}
\tablehead{
\colhead{Sub-Cluster}  & \colhead{Disk Fraction} & \colhead{A}  &  \colhead{R$_{core}$} & \colhead{R$_{tidal}$} & \colhead{AAP$_{meas}$} & \colhead{Number of} & \colhead{$\Sigma$$_{mean}$} & \colhead{$\Sigma$$_{peak}$} \\
\colhead{$ $} & \colhead{$ $} & \colhead{(pc$^{-2}$)$^{1}$} & \colhead{(pc)} & \colhead{(pc)} & \colhead{$ $} & \colhead{members} & \colhead{stars (pc$^{-2}$)} & \colhead{(stars pc$^{-2}$)}}
\startdata
East & 0.32 & 541$\pm$202 & 1.36$\pm$0.30  & 2.32$\pm$0.19 & 1.155 & 809 & 240 & 521 \\
East & 0.39 & 541$\pm$202 & 1.36$\pm$0.30  & 2.32$\pm$0.19 & 1.155 & 664 & 197 & 428 \\
West & 0.44 & 258$\pm$42 & 0.52$\pm$0.11 & 3.1$\pm$1.0 & 1.724 & 501 & 220 & 342 \\
West & 0.53 & 258$\pm$42 & 0.52$\pm$0.11 & 3.1$\pm$1.0 & 1.724 & 416 & 182 & 284 \\
\enddata
       \tablenotetext{1}{This value has not been scaled to correct for the disk fraction.}
\end{deluxetable*}

\newpage 

\begin{deluxetable*}{cccc}
\tablewidth{0pt}
\tablecaption{GAIA DR2 Proper Motions
 \label{tbl:propermotions}}
\tablehead{
\colhead{Sub-Cluster}  & \colhead{$\mu$$_{\alpha}$ (mas yr$^{-1}$)} & \colhead{$\mu$$_{\delta}$ (mas yr$^{-1}$)}  &  \colhead{$\mu$$_{total}$ (mas yr$^{-1}$)}}
\startdata
East & -0.59$\pm$0.02 & -2.32$\pm$0.02 & 2.34$\pm$0.02   \\
West & -1.25$\pm$0.03 & -2.78$\pm$0.02 & 2.86$\pm$0.03   \\
Cep OB3b & -0.69$\pm$0.02 & -2.44$\pm$0.02 & 2.46$\pm$0.02 \\ 
\enddata

\end{deluxetable*}

\end{document}